\documentclass[a4paper,11pt]{article}
\pdfoutput=1
\usepackage{heppub}
\usepackage{amsthm}
\usepackage{amsmath}
\usepackage{bbm}
\usepackage{graphicx,accents}
\usepackage{slashed}
\usepackage{amssymb}
\usepackage{footmisc}
\usepackage{enumitem}
\usepackage{xcolor}
\usepackage{braket}
\usepackage{bm,subcaption}
\usepackage{xfrac}

\newcommand{\bs}{\boldsymbol}

\author[1]{Krishna Rajagopal,}
\author[2]{Bruno Scheihing-Hitschfeld}
\author[1]{and Rachel Steinhorst}

\affiliation[1]{MIT Center for Theoretical Physics -- a Leinweber Institute, Massachusetts Institute of Technology, Cambridge, MA  02139, USA}

\affiliation[2]{Kavli Institute for Theoretical Physics, University of California Santa Barbara, Santa Barbara, California 93106, USA}

\title{Attractors Without Scaling:
Adiabatic Hydrodynamization With and Without Inelastic Scattering}

\preprint{MIT-CTP/5898}

\abstract{We study the process of hydrodynamization in kinetic theories of gluons undergoing boost-invariant expansion using the Adiabatic Hydrodynamization (AH) framework. We study both number-conserving and non-conserving theories, and find that including number non-conserving inelastic scattering processes restores many qualitative features of hydrodynamization in QCD EKT despite the simplicity of our model. In particular, introducing inelastic scattering results in a more realistic hydrodynamization time. With or without number non-conservation, we find that first-order hydrodynamics becomes applicable at the same time that a unique ground state emerges in the dynamical evolution of the one-particle distribution function. Furthermore, we find a set of low-effective-energy attractor modes which evolve adiabatically long before hydrodynamization, and find that the emergence of a gap between these ground state modes and the excited modes coincides with the time at which the system falls onto an attractor surface. Strikingly, this is the case even in the absence of pre-thermal scaling of the gluon distribution function, which has previously been strongly associated with pre-thermal attractor behavior. Finally, motivated by a generic feature we observe in the spectrum, we show that as the system hydrodynamizes, the rapid decoupling of non-hydrodynamic modes
in boost-invariant kinetic theory can be understood with the AH framework in a model-independent fashion.
}

\emailAdd{krishna@mit.edu}
\emailAdd{bscheihi@kitp.ucsb.edu}
\emailAdd{rstein99@mit.edu}

\begin{document}

\maketitle

\section{Introduction}
\label{sec:Intro}

Significant progress has been made in the last decade in developing a microscopic picture of the rapid hydrodynamization of the quark-gluon plasma (QGP) formed in ultrarelativistic heavy ion collisions, both in strongly-coupled holographic descriptions~\cite{Chesler:2008hg,Chesler:2009cy,Chesler:2010bi,Heller:2011ju,Heller:2012je,Heller:2012km,vanderSchee:2012qj,Heller:2013oxa,Casalderrey-Solana:2011dxg,Chesler:2015lsa,Chesler:2015fpa,Heller:2016gbp,Chesler:2015bba,Chesler:2016ceu,Grozdanov:2016zjj,Folkestad:2019lam}, and in weakly-coupled descriptions employing QCD classical field simulations~\cite{PineiroOrioli:2015cpb,Berges:2013fga,Berges:2007re,Berges:2008mr,Berges:2013eia,Gelis:2010nm,Bjoraker:2000cf}, QCD effective kinetic theory (QCD EKT) ~\cite{Baier:2000sb,Mueller:1999fp,Mueller:1999pi,Mueller:2002gd,Arnold:2002zm}, and related kinetic theory descriptions of QCD ~\cite{Kurkela:2014tea,Kurkela:2015qoa,Kurkela:2018vqr,Kurkela:2018wud,Mazeliauskas:2018yef,Boguslavski:2023jvg,Blaizot:2011xf,Blaizot:2013lga,Tanji:2017suk}. In particular, many descriptions of hydrodynamization of QCD matter at both strong and weak coupling have been shown to feature attractor behavior~\cite{Kurkela:2019set,Schlichting:2019abc,Berges:2020fwq}, in the sense that a broad variety of initial conditions rapidly approach (``are attracted to'') the same dynamical evolution.  
Such a loss of memory of the details of initial conditions is a characteristic of hydrodynamization, but at weak coupling this attractor behavior typically arises 
long before hydrodynamics becomes a good description for the system. That is to say, the details of the initial condition become unimportant prior to hydrodynamization. 

In holographic descriptions of hydrodynamization at strong coupling, this attractor behavior is observed during the direct approach of the system to hydrodynamics. However, a distinct type of far-from-equilibrium attractor behavior is observed in weakly-coupled descriptions, distinct from and preceding the attractor behavior seen in the approach to hydrodynamics. This pre-hydrodynamic attractor is typically understood through the lens of a universal self-similarity of the gluon distribution function which has been observed in classical field simulations~\cite{Berges:2013eia,Berges:2013fga} and in kinetic theory~\cite{Tanji:2017suk,Mazeliauskas:2018yef}. That is, long before hydrodynamics, the gluon distribution function takes the scaling form 
\begin{equation}\label{eq:scalingform}
    f_s(p_\perp,p_z,\tau) = \tau^{\alpha} w_s\left( p_\perp \tau^{\beta}, p_z \tau^{\gamma}\right)\,,
\end{equation}
in which the scaling exponents $\alpha,~\beta,$ and $\gamma$ as well as the underlying shape of $w_s$ are universal in the sense that they do not depend on details of the initial conditions. 
QCD kinetic theory has long been known to exhibit scaling behavior
%{\color{magenta} BSH: there is no reference to a scaling solution per se in BMSS, so if BMSS is all we cite in this sentence, I prefer to say scaling behavior} %a scaling solution 
(referred to as BMSS scaling) at very early times when the gluon distribution function is over-occupied, with scaling exponents $\alpha=-2/3$, $\beta=0$, and $\gamma=1/3$~\cite{Baier:2000sb}.
More recently,  the scaling solution in the underoccupied regime (that arises subsequent to BMSS scaling) has been referred to as ``dilute" scaling, with $\alpha=-1$ and $\beta=\gamma=0$~\cite{Brewer:2022vkq}.
As the distribution function later approaches hydrodynamics, it continues to take this form, with $w_s$ becoming a Boltzmann (or Bose-Einstein or Fermi-Dirac) distribution and the scaling exponents taking on the ``thermal'' values $\alpha=0$ and $\beta=\gamma=1/3$. 
%However, QCD kinetic theory has long been known to exhibit a second scaling solution in the overoccupied regime called ``BMSS" scaling ($\alpha=-2/3,~\beta=0,~\gamma=1/3$)~\cite{Baier:2000sb}, and a similar phenomenon in the underoccupied regime has recently been referred to as ``dilute" scaling ($\alpha=-1,~\beta=\gamma=0$)~\cite{Brewer:2022vkq}. 
Far-from-equilibrium attractor behavior is distinct from the ``thermal" attractor because it is not directly associated with the approach to hydrodynamics, and can even occur in simplified systems which do not hydrodynamize. 

The Adiabatic Hydrodynamization (AH) picture provides a framework for understanding why, how, and when each type of attractor behavior occurs as well as the relations between different attractors. It does so by casting kinetic theory 
in the form $\partial_t|\psi\rangle=-H_{\rm eff}(t)|\psi\rangle$ (here schematic; described concretely in Sections~\ref{sec:adiabatic_hydrodynamization} and~\ref{sec:results})  and positing
that each attractor is described as the evolving instantaneous ground state of the effective Hamiltonian $H_{\rm eff}(t)$ during an epoch in which excited states are well separated from the ground state and the time evolution satisfies the adiabatic condition. The logic behind the adiabatic condition carries over from quantum mechanics, but the replacement of the $i$ 
in quantum evolution by a negative sign means that the time evolution here differs in an important way:
it implies that the occupations of all
states whose eigenvalue of $H_{\rm eff}$
are above the lowest eigenvalue will decay, making the lowest eigenvalue state(s) an attractor.
 
This was first proposed in Ref.~\cite{Brewer:2019oha}, and has been demonstrated in simplified QCD kinetic theories in Refs.~\cite{Brewer:2022vkq,Rajagopal:2024lou}, with Ref.~\cite{Rajagopal:2024lou} providing explicit examples of how $H_{\rm eff}(t)$ can describe the non-adiabatic time evolution as the system transitions from one attractor to the next. 
The authors of Ref.~\cite{Brewer:2022vkq} employed the Boltzmann equation in the small-angle scattering approximation~\cite{Mueller:1999pi,Blaizot:2013lga} at early times in a boost-invariant weakly coupled plasma, obtaining a simplified theory in which the analytic form of the pre-thermal attractor is known explicitly and captures the scaling exponents of the earliest epoch of the BMSS ``bottom-up'' scenario~\cite{Baier:2000sb}. However, because of the simplifying assumptions made at early-time in Ref.~\cite{Brewer:2022vkq}, this simplified theory does not reach local thermal equilibrium.
In the latter study~\cite{Rajagopal:2024lou}, although some key approximations made in Ref.~\cite{Brewer:2022vkq} --- such as neglecting gluon number changing processes --- are maintained,  we were able to relax other simplifying assumptions and obtain a unified description of the entire hydrodynamization process in a kinetic theory of gluons with elastic small-angle scatterings,
%presents a closely related theory which hydrodynamizes, 
smoothly connecting the pre-thermal attractor behavior described by $H_{\rm eff}(t)$ with a band of ground states to the later thermal attractor through a rapid (non-adiabiatic) evolution in which a single ground state emerges. 
Because this analysis neglects inelastic $1\rightarrow 2$ processes, hydrodynamization takes much longer than it would in QCD EKT. This provides an important motivation for the work that we shall report here, in which for the first time we employ AH to analyze hydrodynamization in kinetic theory including inelastic processes.

Our AH framework analysis in Ref.~\cite{Rajagopal:2024lou} revealed a crucial qualitative feature of the hydrodynamization process: that each ``attractor'' phase is characterized by the time-dependent ground state(s) of the time evolution operator $H_{\rm eff}$ of the (appropriately rescaled) distribution function, and that in this picture each one of these phases is reached dynamically via the decay of the ``excited'' states of this operator. Loss of memory of the initial condition is understood in this way, with the only remaining information encoded in the coefficient(s) of the ground state(s). In practice, we found that the early-time attractor phase is characterized by a low-energy ``band,'' whereas the late-time, hydrodynamizing attractor phase is characterized by a unique ground state, which can be identified as a single surviving linear combination of the states in the pre-thermal attractor band after the eigenvalues of all the other states separate from that of this survivor. Importantly, the characterization of the early-time attractor by a band of eigenstates rather than a single solution differs from the traditional notion of an attractor in that the initial-state memory associated with the excited eigenstates is forgotten, but the system does still retain some memory of the initial state in that the occupation of various states within the ground state band may be different. We can think of this as an ``attractor surface" rather than a single attractor, and systems with different initial conditions may fall to different linear combinations of solutions within the surface. For a scaling solution as in Eq.~\eqref{eq:scalingform}, an attractor surface rather than a traditional attractor implies that $w_s$ is restricted in its form, but is not universal. This may, but need not, result in differing evolution of physical quantities such as anisotropy, occupancy, energy density or number density during the pre-thermal attractor regime. (This aspect of such an attractor surface was not explored in Ref.~\cite{Rajagopal:2024lou}; we shall explore it in Sec.~\ref{sec:results}.) In summary, we established in Ref.~\cite{Rajagopal:2024lou} that in this kinetic theory the memory of the initial condition is lost sequentially:
the memories lost during the very early approach to the prehydrodynamic attractor surface are never regained; the subsequent transition from the pre-thermal attractor surface to the hydrodynamizing attractor only involves further loss of memory.
The AH framework provides a unified description of this entire process, in both the pre-thermal and hydrodynamizing stages~\cite{Rajagopal:2024lou}.

However, one significant shortcoming of this previous work is the omission of number non-conserving processes, that is, inelastic scatterings. 
In the bottom-up picture, inelastic $1 \leftrightarrow 2$ processes are key to rapid hydrodynamization. Indeed, the hydrodynamization times found in Ref.~\cite{Rajagopal:2024lou} are exponentially longer than is realistic, even with physically reasonable couplings. 
Furthermore, since hydrodynamization with inelastic scatterings included will occur by a different mechanism (the formation of a thermal bath of soft gluons), it is important to determine whether the adiabatic picture of attractors during hydrodynamization continues to hold once we account for inelastic scatterings. In this work we answer this question.  We shall see that the AH framework {\it does} describe hydrodynamization in this setting --- whether or not the evolution of the distribution function takes the scaling form (\ref{eq:scalingform}).

We find that, as expected, including inelastic scattering processes leads to much earlier hydrodynamization, and that the acceleration of hydrodynamization by $1\rightarrow 2$ processes is most dramatic for the weakest couplings we are able to evolve to hydrodynamization. 
With a realistic (intermediate) value of the coupling, upon including inelastic processes we find that hydrodynamization occurs on a realistic timescale.
Furthermore, incorporating inelastic scattering leads to a more QCD EKT-like evolution of the system in the  anisotropy vs. occupancy plane~\cite{Kurkela:2015qoa,Boguslavski:2023jvg}. When we introduce number non-conservation, we do indeed find the expected ``hook-shaped'' turn back towards higher occupancies as the system isotropizes. We also find that either with or without inelastic scatterings, there appears to be a pre-thermal attractor surface reached long before the onset of hydrodynamics, even in the absence of scaling behavior. 
We show evidence that the attractor behavior can be explained by the AH hypothesis:  whether or not scaling behavior is seen, each attractor is described by the lowest-eigenvalue  eigenstate or eigenstates of $H_{\rm eff}$, separated from higher-eigenvalue states by a gap sufficiently large that any non-attractor modes decay away exponentially in log time (that is, power law decay towards the attractor in linear time). This is a new --- and more general --- picture of pre-thermal far-from-equilibrium attractors in the initial stages of QGP formation, which were previously understood only through the presence of scaling. We show that while scaling can characterize pre-thermal attractors, it is not a necessary condition. And, we show that AH can describe attractors even in the absence of scaling behavior.

Our approach is not (yet) the final word on how QCD EKT hydrodynamizes, because we have not yet included quarks in the kinetic theory or initial conditions with spatial inhomogenities, as in realistic heavy ion collisions. 
Our work is nevertheless a significant milestone in establishing how hydrodynamization takes place in a theory of massless gluons with both elastic and inelastic scatterings, establishing that the loss of memory of the initial condition is sequential, and showing that the different attractor phases in the system's evolution need not exhibit any characteristic scaling pattern. Rather, they are characterized by the instantaneous spectrum of the effective Hamiltonian of the theory -- the central object in the AH framework -- with each attractor regime arising when energy gaps emerge.

The rest of this paper is organized as follows. In Section \ref{sec:adiabatic_hydrodynamization}, we provide a brief introduction to the AH framework. Section \ref{sec:results} contains the main results of this paper, first outlining our methods and then discussing in turn pre-thermal attractor behavior and the subsequent hydrodynamizing attractor at very weak and somewhat stronger couplings, demonstrating that the AH framework provides a unified description of these sequential attractors and hydrodynamization whether or not the kinetic theory exhibits scaling.  Finally, in Section \ref{sec:hydro_inevitability} 
we focus on a regularity seen in the late-time behavior of the eigenvalues of $H_{\rm eff}$ in all of our results, and show that this behavior holds for a broad class of kinetic theories (which we specify) that describe boost-invariant longitudinal expansion. In this context, the AH framework  yields an intuitive and generic understanding of how all of the non-hydrodynamic modes in a kinetic theory decouple at late times, after hydrodynamization. We give our concluding remarks and look ahead in Section~\ref{sec:outlook}.

\section{The Adiabatic Hydrodynamization Framework}\label{sec:adiabatic_hydrodynamization}

We begin our discussion by introducing the AH framework in a general setting.
Consider an evolution equation of the form
\begin{equation}\label{eq:se}
    \frac{d\psi}{dt}=-H[\psi](t)\psi(q_1,q_2,\ldots q_d)\,.
\end{equation}
The applicability of the AH framework as we will describe it in this Section is not limited to kinetic theories; we shall see by explicit construction in the next Section, though, that within kinetic theory 
the Boltzmann equation can be rearranged to take the form \eqref{eq:se}. In this context,
 $\psi$ is the distribution function $f$ that depends on the components $q_i$ of the particle momenta in $d$ dimensions, and the effective Hamiltonian-like operator $H[f]$ (henceforth we drop the subscript eff) will contain both the collision kernel $C[f]$ and expansion terms.
 The action of $H[\psi](t)$ on $\psi$ at a fixed value of its arguments $\psi$ and $t$ is linear. However, given that the evolution equation has $\psi$ as an argument of $H$, the ensuing dynamics is in general nonlinear. The operator $H[\psi](t)$ can in general be non-Hermitian.  For this reason, as well as because the prefactor on the RHS of Eq.~(\ref{eq:se}) does not include an $i$, this evolution is not the same as that in quantum mechanics. 

We may nonetheless consider an analogy to quantum mechanics and to the adiabatic theorem, in the sense that if $H$ is gapped and sufficiently slowly-varying, by the same logic as in the quantum mechanics case, if $\psi$ is initially in the lowest-eigenvalue state of $H$ it will remain there under time evolution. Furthermore, because of the $-1$ in 
Eq.~\eqref{eq:se} in place of the $i$ which appears in the Schr\"{o}dinger equation, excited states with eigenvalue $\epsilon (t)$ evolve proportionally to $e^{-\epsilon(t) \delta t}$ rather than the usual $e^{-i\epsilon(t) \delta t}$, meaning that the occupation of 
all states with larger eigenvalues (``excited states'') 
will die off if there is a gap between their eigenvalues and the lowest eigenvalue of the effective Hamiltonian $H$ and the evolution is adiabatic. Thus, regardless of what the initial condition is for the kinetic theory, which is to say regardless of the initial occupations of the eigenstates of $H$, if there is a gap between one or more lowest eigenvalue states and all the higher eigenvalue states and if the time evolution is adiabatic the resulting evolution will rapidly be attracted to the lowest eigenvalue state or states --- the ground state(s) of $H$ --- after sufficient time has passed. That is, the ground state (or band of ground states) of $H$ acts as an attractor (or attractor surface). If the ground state is degenerate or nearly degenerate, different initial conditions may fall onto different solutions on the resulting attractor surface, and subsequent time evolution may take the system from one linear combination of solutions on the attractor surface to another. 
All the solutions on the attractor surface 
may, but need not, share the same physical characteristics such as energy density, pressure, anisotropy, and occupancy. If this is the case, then in terms of these characteristics the collapse of varied initial states onto the attractor surface looks just as universal as attraction to a single attractor.

An evolution equation such as a Boltzmann equation can always be cast into the form of Eq.~\eqref{eq:se}, but in general the naive way of doing this will not result in a time-varying effective Hamiltonian with a gapped, slowly-varying, spectrum. (In fact, a broad class of collision kernels are gapless \cite{Gavassino:2024rck}.) The idea behind AH is to take an arbitrary equation of the form 
of Eq.~\eqref{eq:se} and perform some appropriate time-dependent coordinate rescaling
\begin{equation} \label{eq:rescaling_example}
    \psi(q_1,q_2,\ldots, q_d)=C_0(t) \psi '\left(\frac{q_1}{C_1(t)},\frac{q_2}{C_2(t)},\ldots,\frac{q_d}{C_d(t)},t \right)
\end{equation}
such that the evolution equation for the rescaled function $\psi'$
\begin{equation}
    \frac{d\psi'}{dt}=-H'[\psi'](t)\psi'(q_1',q_2',\ldots 
    q_d',t)\,,
\end{equation}
which we will later derive from the Boltzmann equation, is adiabatic. Thus, 
after sufficient time the state $\psi'$ can be described by an underlying time-dependent final state that subsequently evolves adiabatically together with the time-dependent coordinate rescalings $C_0(t)$, $C_1(t)$, $\ldots$, $C_d(t)$. Here, the final state of $\psi'$ will in general be a linear combination of the states in the ground state band of $H'$.
If $H'$ has only a single lowest eigenvalue state, the final state of $\psi'$ is this state.
This represents a significant loss of memory of the initial condition of the kinetic theory distribution function,
which is to say it describes an attractor.

Any circumstance in which both attractor behavior and time-dependent scaling 
as described by the universal scaling form Eq.~\eqref{eq:scalingform} 
have been demonstrated in a kinetic theory description of a gas of gluons
is a natural context in which to look for an adiabatic interpretation by performing a coordinate rescaling as described above~\cite{Brewer:2022vkq}. This is because when $f$ is scaling, the time-dependent factors associated with scaling exponents $\alpha$, $\beta$, and $\gamma$ are natural candidates for the coordinate rescalings $C_i(t)$ described above.
%Here we will consider a kinetic theory description of a gas of gluons. 
In far-from-equilibrium descriptions of a boost-invariant, longitudinally expanding gas of gluons, there are several known examples of scaling behavior of the type in Eq.~\eqref{eq:scalingform} which may arise.
For example, free-streaming corresponds to $(\alpha,\beta,\gamma)=(0,0,1)$, although in this case the underlying rescaled distribution $w_s$ does not act as an attractor meaning that this case is not universal. 
In the ``bottom-up'' picture of hydrodynamization~\cite{Baier:2000sb}, an initially overoccupied distribution of primarily hard gluons evolves according to Eq.~\eqref{eq:scalingform} with $(\alpha,\beta,\gamma)=(-2/3,0,1/3)$ --- the BMSS fixed point. This describes a broadening in $p_z$ relative to free-streaming due to elastic scatterings, which is enhanced due to the Bose factor in the density of scatterers. When the occupation drops below 1 and Bose enhancement becomes unimportant (or when the Bose factor is omitted, as in the kernel we will study below), we instead have $(\alpha,\beta,\gamma)=(-1,0,0)$ --- the ``dilute" fixed point~\cite{Brewer:2022vkq}. Both the BMSS and dilute fixed points act as pre-thermal attractors with $w_s$ approximately Gaussian, as shown first for the BMSS fixed point in Ref.~\cite{Berges:2013fga,Berges:2013eia}, and later for both in Ref.~\cite{Brewer:2022vkq}.
When inelastic $1\rightarrow 2$ processes are included, emissions from the hard gluons populate the soft sector, leading to the bottom-up stage of bottom-up hydrodynamization. 
The temperature characterizing the soft sector gluon distribution increases as $1\rightarrow 2$ processes pump energy in from from the hard sector until a time  $\tau \sim \alpha^{-13/5} Q_s^{-1}$, at which time the system has thermalized, with $w_s$ having reached the  Bose-Einstein distribution~\cite{Baier:2000sb}.  The subsequent time evolution is described by hydrodynamics with boost-invariant longitudinal expansion
and
thermal scaling exponents $(\alpha,\beta,\gamma)=(0,1/3,1/3)$. 
(While it is possible for thermalization to occur with elastic collisions only, this is  not  ``bottom-up" thermalization in the sense of populating a thermal bath of soft gluons, and will take place on a significantly longer time scale, if at all.) We also stress that although these examples as stated have constant scaling exponents, in general $\alpha,~\beta,~\gamma$ can evolve in time, and $f$ can take the scaling form \eqref{eq:scalingform} already before $\alpha,~\beta,~\gamma$ approach the constant values associated with each attractor regime~\cite{Mazeliauskas:2018yef}. This provides ample motivation for an adiabatic interpretation of all stages of hydrodynamization, although scaling is neither a necessary nor sufficient condition for AH to hold as an explanation for and predictor of attractor behavior.

In any case in which $w_s$ as in Eq.~\eqref{eq:scalingform} is an attractor, we would like to identify the scalings 
$C_i$ which appear in Eq.~\eqref{eq:rescaling_example} and relate them to the scaling exponents
$\alpha$, $\beta$, and $\gamma$. 
If the resulting effective Hamiltonian $H'$ describing the evolution of $w$ has time-independent eigenvectors, then these eigenstates describe exact scaling solutions.
Unlike the free-streaming case, each of the three scaling regimes associated with bottom-up hydrodynamization (BMSS, dilute and thermal) is characterized not only by scaling, but by a universal underlying distribution function $w_s$ which acts as an attractor. 
(The underlying $w_s$ of free-streaming is non-universal and does not act as an attractor.) 
It is natural to understand this attractor behavior through the lens of AH, in which $w_s$ is the ground state distribution of an effective Hamiltonian
\begin{equation}\label{eq:Hw=-partialw}
    H w(\zeta,\xi,\tau) = -\partial_y w(\zeta,\xi,\tau)
\end{equation}
where $\zeta\equiv p_\perp \tau^{-\beta}$ and $\xi \equiv p_z \tau^{-\gamma}$. This adiabatic picture has been demonstrated analytically~\cite{Brewer:2022vkq} for the BMSS and dilute fixed points in a simplified version of kinetic theory with only small-angle elastic scattering that does not thermalize. In this case, the effective Hamiltonian $H$ has eigenvalues
\begin{equation}\label{eq:bsy_energies}
    \varepsilon_n = 2n(1-\gamma) -2m\beta\,
\end{equation}
labeled by non-negative integers $n$ and $m$. (Here, the eigenstates of $H$ are products of a mode function of the transverse momenta that depends on $m$ and a mode function of the longitudinal momentum that depends on $n$.) For constant values of $\alpha$, $\beta$, and $\gamma$, this result predicts power law decay of the occupation of all excited state modes---a consequence of the fact that the prefactor on the RHS of Eq.~\eqref{eq:Hw=-partialw} is a minus sign, not an $i$. 
(This power law decay was later noted and expanded upon 
in Ref.~\cite{DeLescluze:2025jqx}.)
In previous work~\cite{Rajagopal:2024lou}, we have furthermore numerically demonstrated that in a kinetic theory with a simplified small-angle scattering kernel which \textit{does} thermalize:
\begin{enumerate}
    \item The pre-thermal effective energy levels are still approximately described by~\eqref{eq:bsy_energies} with $\beta\approx 0$, meaning that pre-thermal attractors are not a unique ground state, but an attractor surface consisting of an effectively degenerate set of transverse modes. In the BMSS and dilute stages, the system loses ``longitudinal memory". That is to say in the language of Eq.~\eqref{eq:bsy_energies}, when $\beta\approx 0$, changing the transverse mode by incrementing $m$ hardly changes the energy, staying within an
    approximately degenerate band of eigenstates, while incrementing $n$ and changing the longitudinal mode jumps from one band of modes of the effective Hamiltonian to another. Consequently, occupation of eigenstates with excited longitudinal degrees of freedom (i.e. large $n$) quickly die away as suggested by the AH hypothesis, leaving only occupation of the attractor surface made up of eigenstates of differing $m$. 
    
    \item The thermal attractor is a unique ground state, and emerges from the pre-thermal ``ground state band" of modes with no mixing from the excited longitudinal modes. By this we mean that the system transitions from the pre-thermal energy levels roughly characterized by Eq.~\eqref{eq:bsy_energies}, in which for each $m$ there is an energy level near zero, to a single, non-degenerate, effective energy level at 0. All other effective energies increase exponentially at late times, a general phenomenon that we shall explain in Sec.~\ref{sec:hydro_inevitability}, and therefore the gap increases exponentially. This means that the single ground state mode, which corresponds to the equilibrium distribution, becomes more and more attractive over time. Importantly, the lack of mixing means that the equilibrium solution that at late times becomes the single ground state of $H$ lies entirely within the earlier pre-hydrodynamic attractor surface, and all memory of the initial conditions that was lost in the pre-thermal phase remains lost as the system falls onto the hydrodynamizing attractor.
\end{enumerate}
That is, memory loss occurs irreversibly in two distinct and progressive stages. However, the work of Ref.~\cite{Rajagopal:2024lou} was restricted to a kinetic theory with elastic scatterings only, which means that it could not describe the full bottom-up picture.
In this work, we rectify the omission.

\section{Hydrodynamization With and Without Inelastic Scattering}\label{sec:results}

We will use a simplified version of QCD effective kinetic theory to describe the evolution of the gluon distribution function $f$. Assuming boost invariance and translational invariance, the Boltzmann equation takes the form
\begin{equation} \label{eq:boltzmanneqn}
    \frac{\partial f}{\partial \tau} - \frac{p_z}{\tau} \frac{\partial f}{\partial p_z} = - C^{2\leftrightarrow 2}[f]-C^{1\leftrightarrow 2}[f] \, ,
\end{equation}
where $C^{2\leftrightarrow 2}[f]$ and $C^{1\leftrightarrow 2}[f]$ are the collision kernels for elastic and inelastic scatterings respectively. This is the evolution equation that we shall cast in the form of an effective Hamiltonian evolution equation \eqref{eq:se} for a rescaling of $f$.

The AH framework has previously been applied to kinetic theories with only elastic scatterings, in the small-angle-scattering limit~\cite{Mueller:1999pi,Serreau:2001xq,Blaizot:2013lga}, in which the collision kernel takes the form
\begin{equation}
	\label{eq:small-angle-kernel}
  \mathcal{C}^{2\leftrightarrow 2}[f] = -\lambda_0 \ell_{\rm Cb} [f] \left[ I_a[f] \nabla_{\bf p}^2 f + I_b[f] \nabla_{\bf p} \cdot \left( \frac{\bf p}{p} (1+f) f \right) \right]\, ,
\end{equation}
where $\lambda_0 \equiv 4\pi \alpha_s^2 N_c^2=4\pi \lambda_{\rm 't~Hooft}^2$, with the functionals $I_a$, $I_b$, and $\ell_{\rm Cb} [f]$ given by
\begin{align}
  I_a[f] = \int_{\bf p} f (1+f), \, 
  \qquad
  I_b[f] = \int_{\bf p}\, \frac{2}{p} f,\, 
  \qquad
  \ell_{\rm Cb} [f] =\ln \left( \frac{p_{\rm UV}}{p_{\rm IR}} \right)\, ,
\end{align}
using the shorthand notation $\int_{\bf p} \equiv \int \frac{d^3 p}{(2\pi)^3}$.
The integral $I_{a}$ represents the effective density of scatterers, where the factor $(1+f)$ is due to Bose enhancement of the gluons.  In thermal equilibrium, $I_b=I_a/T$ 
and so even out of equilibrium we shall refer to an effective temperature $T_{\rm eff}$ with  $1/T_{\rm eff}\equiv I_b/I_a=\braket{2/p}$, where $\braket{\cdot}$ means $(\int_{\bm{p}} \cdot\; f)/(\int_{\bm{p}} f)$. $I_{b}$ is also related to the Debye mass $m_{D}$ through
\begin{equation}
\label{mD}
  m_D^2 = 2 N_c g^2_{s} I_b\, ,
\end{equation}
as $I_b=T^2/(2N_c)$ for gluons with a Bose-Einstein distribution.
The Coulomb logarithm $\ell_{\rm Cb} [f]$ is a perturbatively divergent integral over the small scattering angle~\cite{Mueller:1999pi}, which (following Refs.~\cite{Mueller:1999fp,Brewer:2022vkq,Rajagopal:2024lou}) we evaluate as $\ell_{\rm Cb} [f]=\ln (p_{\rm UV}/p_{\rm IR})$ with $p_{\rm IR} = m_D$ and $p_{\rm UV} = \sqrt{\langle p^2 \rangle}$.

Without inelastic scatterings, as has long been known, hydrodynamization occurs on a significantly longer timescale than when inelastic processes are included in the kinetic theory and than in heavy ion collisions. In this paper, we incorporate inelastic $1\rightarrow 2$ processes, albeit employing a simplified form for the inelastic piece of the collision kernel to which we can more easily apply AH. We start from the form found in Ref.~\cite{Baier:2000sb}, in which the inelastic part of the collision kernel is
\begin{align}
    C^{1\leftrightarrow 2}[f] &= -\int_0^1 dx \frac{d^2 I}{dx\,dt} \left\{ \frac{1}{x^{5/2}} \left[ f\left(\frac{p}{x}\right)(1+f(p))\left( 1+ f\left( \frac{p(1-x)}{x}\right) \right)\right.\right.\nonumber\\
    &\qquad\qquad\qquad\qquad\qquad\qquad\qquad \left.-f(p) f\left( \frac{p(1-x)}{x}\right) \left( 1+ f\left(\frac{p}{x}\right)\right) \right]\\
    &\qquad \left. -\frac{1}{2} \left[ f(p)(1+f(px))(1+f(p(1-x)))-f(px)f(p(1-x))(1+f(p)) \right]
    \right\}\,.\nonumber
\end{align}
where
\begin{equation}
    \frac{d^2I}{dx\,dt}=\lambda_0 \sqrt{\frac{I_a[f] \ell_{\rm Cb} [f]}{2\pi^3 p}}\frac{(1-x+x^2)^{5/2}}{(x-x^2)^{3/2}}
\end{equation}
represents the rate of nearly collinear splitting for a hard gluon.
We can write $\varepsilon\equiv 1-x$ and expand the integrand about $\varepsilon =0$, which is essentially the limit of soft splitting. This expansion will significantly underestimate the collision kernel for small $p$, but for $p\gtrsim T_{\rm eff}$ it is a reasonable approximation for non-pathological $f(p)$ (i.e., as long as the expansion %KR: WHY VARIOUS EXPANSIONS? THIS PARENTHETICAL INSERTION IS OK WITH ME IF WE JUST SAY THE EXPANSION INSTEAD OF THE VARIOUS EXPANSIONS. 
of $f$ in terms of $\varepsilon$ is under control). If we perform the expansion to ${\cal O}(\varepsilon)$ and keep those ${\cal O}(\varepsilon^2)$ terms that are needed in order to ensure that the collision kernel still conserves energy,
we have
\begin{align}\label{eq:inelastic-kernel}
    C^{1\leftrightarrow 2}[f]&=-2\lambda_0 \sqrt{\frac{I_a[f] \ell_{\rm Cb} [f]}{2\pi^3 p}} \left[ \frac{g}{p} \left( \bm{p}\cdot \nabla_{\bm{p}} \right)^2 +(1+2f) \left( \bm{p}\cdot \nabla_{\bm{p}} \right)\right. \nonumber\\
    &\qquad\qquad\qquad\qquad\qquad\qquad\qquad\left.+\frac{7}{2} \left( 1+f+\frac{g}{p}\left( \bm{p}\cdot \nabla_{\bm{p}} \right)\right)\right]f
\end{align}
where
\begin{align}
    g\equiv \lim_{p\rightarrow 0} p f(\bm{p})\ .
\end{align}
This is convenient in the sense that we are easily able to write $C^{1\leftrightarrow 2}[f]$ as a (comparatively) simple operator acting on $f$. However, including explicit factors of $f$ in the Hamiltonian causes numerical challenges in the AH framework. Omitting Bose enhancement (such that the kernel is solved by a Boltzmann distribution rather than a Bose-Einstein distribution), we instead have
\begin{align}\label{eq:1to2final}
    C^{1\leftrightarrow 2}[f]=-2\lambda_0 \sqrt{\frac{I_a[f] \ell_{\rm Cb} [f]}{2\pi^3 p}} (1-f(\bm{p}=0))\left( \frac{7}{2} +\bm{p}\cdot \nabla_{\bm{p}} \right) f\,,
\end{align}
which is the form we use to obtain the results of this paper. Accordingly, we will also omit the Bose enhancement terms from the elastic kernel~\eqref{eq:small-angle-kernel}, using
\begin{equation}
	\label{eq:2to2final}
  \mathcal{C}^{2\leftrightarrow 2}[f] = -\lambda_0 \ell_{\rm Cb} [f] \left[ I_a[f] \nabla_{\bf p}^2 f + I_b[f] \nabla_{\bf p} \cdot \left( \frac{\bf p}{p} f \right) \right]\, ,
\end{equation}
and will also omit the factor of $(1+f)$ from $I_a[f]$.

We expect on general physical grounds that including inelastic $1\rightarrow 2$ processes in the kinetic theory will speed up the process of hydrodynamization, as it means that high momentum particles can share their momentum among more and more softer particles. One might wonder whether our simplified $1\leftrightarrow 2$ kernel still encapsulates the physics necessary for this speedup. We first note that the inelastic collision kernel  does still contain communication between the hard and soft sectors through the presence of 
$f({\bf p}=0)$ in the expression \eqref{eq:1to2final} at any momentum. The fact that this appears as a factor $1-f({\bf p}=0)$ furthermore 
fixes $f({\bf p}=0)=1$ at late times, and in so doing fixes the relation between the late-time Debye mass and the (effective) temperature. In contrast, in the absence of inelastic processes the number density of gluons $n$ is strictly proportional to $1/\tau$
in a boost invariant longitudinally expanding plasma and $f(\bf{p}=0)$ is sensitive to the initial state.
Additionally, we can note that without Bose corrections, $I_a=n$,
so $C^{2\leftrightarrow 2}[f]\sim{n}$ while $C^{1\leftrightarrow 2}[f]\sim \sqrt{n}$. 
Once $n$  is decreasing approximately as $1/\tau$, 
therefore,
$C^{1\leftrightarrow 2}[f]/C^{2\leftrightarrow 2}[f]\sim \sqrt{\tau}$, meaning that $C^{1\leftrightarrow 2}[f]$ stays larger for longer, and thus we might guess that hydrodynamization will happen more quickly with its inclusion (despite our simplification of the inelastic kernel). 
We will see that this is indeed the case. 

We can also compare these expectations to the bottom-up description of what inelastic collisions ought to do in the process of thermalization~\cite{Baier:2000sb}. BMSS describe this process in two stages: first, relatively soft emissions from the hard gluons populate the soft sector. Then, the radiated gluons undergo hard branchings which cascade into further hard branchings, forming the thermal bath. Inspecting our soft limit, clearly we have removed the second stage: no hard branchings will occur. Nonetheless, the flow of energy to the soft sector will still occur, with thermalization proceeding through soft inelastic scatterings and elastic scatterings.

In general, $H$ is time-dependent and
an important part of AH is choosing an appropriate time-dependent basis, and appropriate time-dependent coordinate rescalings, such that the time-evolution of the rescaled $H$ is as adiabatic as possible. In Sec.~\ref{sec:ahsetup}, we will describe our basis choice and prescription for time-dependent scalings for the AH analysis of the kinetic theory specified by Eqs.~\eqref{eq:boltzmanneqn}, \eqref{eq:1to2final} and \eqref{eq:2to2final}. Then, in the following two sections we will present results and analyze how the inclusion of $1 \leftrightarrow 2$ processes affects the adiabatic interpretation of prehydrodynamic attractor behavior (Sec.~\ref{sec:weakcoupling}) and hydrodynamization (Sec.~\ref{sec:strongcoupling}).

\subsection{Choice of Basis and Scaling}\label{sec:ahsetup}

We will make the same choice of basis and scaling as in the latter part of Ref.~\cite{Rajagopal:2024lou}, which we describe briefly here. 
Because $H$ is not Hermitian, we choose different bases for functions on which $H$ acts from the left and right.
Anticipating that the early-time attractor surface consists of modes that are approximately Gaussian in $p_z$ (see Ref.~\cite{Brewer:2022vkq}) and that the late-time attractor is approximately thermal, we choose the right and left bases 
\begin{align}\label{eq:choice-of-basis}
    \psi_{nl}^{(R)} = N_{nl} e^{-\chi} e^{-u^2 r^2/2} L_{n-1}^{(2)}(\chi) Q_l^{(R)}(u;r) \, , & & \psi_{nl}^{(L)} = N_{nl} L_{n-1}^{(2)}(\chi) Q_l^{(L)}(u;r) \, ,
\end{align}
where $\chi=p/D(y)$ is the rescaled overall momentum with some time-dependent scaling $D(y)$, and $u\equiv p_z/p$. The log-time $y\equiv \ln(\tau/\tau_I)$ refers to an initialization time $\tau_I$; we define our units of time such that $\tau_I=1$. (In a heavy ion collision, $\tau_I$ would be the earliest time at which a kinetic theory description is possible, often taken to be $\sim 0.1$~fm$/c$.) The polynomials $L_{n-1}^{(2)}(\chi)$ and $Q_l^{(R,L)}(u;r)$ where $n$ are positive integers and $l$ are non-negative integers are chosen to make the basis states orthogonal, which in the case of $L_{n-1}^{(2)}(\chi)$ means that they are generalized Laguerre polynomials. The polynomials $Q_l^{(R)}(u;r)$ and $Q_l^{(L)}(u;r)$, which only differ by a multiplicative constant that is a function of $r$ and not $u$, are proportional to Legendre polynomials in the limit $r \to 0$, and proportional to Hermite polynomials in the limit $r \to \infty$. The constants $N_{nl}$ are normalization constants. The parameter $r$ characterizes the degree of anisotropy of the basis (with $r=0$ corresponding to $l =0$ basis states that are isotropic in momentum space). We choose $r$ to be time-dependent, namely $r(y)$, with $r$ chosen to reflect the initial conditions at early times and $r\rightarrow 0$ at late times. We specify our choice of $r(y)$ explicitly below.
The choice of basis \eqref{eq:choice-of-basis} has the feature that for $r\rightarrow 0$, the first basis state $\psi_{10}^{(R)}$ is a Boltzmann distribution, while for $r\rightarrow \infty$, it is approximately Gaussian along $u$.  We will in all cases discussed here use a total of 12 basis states, the states for which $l$ is even and $n+l\leq 6$. The states with odd $l$ are not relevant because we assume $f$ is even in $p_z$.

We write
\begin{equation}\label{eq:A-w-definition}
f(p,u,y)=A(y) w(\chi,u,r(y),y) 
\end{equation}
for $f$ evolving according to Eq.~\eqref{eq:boltzmanneqn} with $C^{1\leftrightarrow 2}[f]$ and $C^{2\leftrightarrow 2}[f]$ as in Eq.~\eqref{eq:1to2final} and~\eqref{eq:2to2final}, respectively.
Defining the effective Hamiltonian $H$ according to
\begin{equation}\label{eq:Schrodinger-like}
    \partial_y w = -Hw\,,
\end{equation}
we can derive the explicit expression for the $H$ that governs this kinetic theory: 
\begin{align}\label{eq:hamiltonian}
    H &= \frac{\partial_y A}{A} + (\partial_y r) \partial_r -\frac{\partial_y D}{D} \chi \partial_\chi - u^2 \chi \partial_\chi - u(1-u^2) \partial_u - \tau \lambda_0 \ell_{\rm Cb} \frac{I_b}{D} \left[ \frac{2}{\chi} + \partial_\chi \right] \nonumber \\ 
    & \quad - \tau \lambda_0 \ell_{\rm Cb} \frac{I_a}{D^2}  \left[ \frac{2}{\chi} \partial_\chi + \partial_\chi^2 + \frac{1}{\chi^2} \left( (1-u^2)\frac{\partial^2}{\partial u^2} -2u \frac{\partial}{\partial u} \right) \right] \nonumber \\ 
    & \quad - 2\tau\lambda_0\sqrt{\frac{\ell_{\rm Cb}I_a}{D\pi^3}}\left(1-A\,w\Big|_{\chi\rightarrow 0}\right)\left( \frac{7}{2}\frac{1}{\sqrt{\chi}}+\sqrt{\chi}\,\partial_\chi\right)\,.
\end{align}
(Note that in our basis the $\chi\rightarrow 0$ limit of $w$ that appears in \eqref{eq:hamiltonian} depends on $u$, and in practice we calculate it by averaging over $u$.)
Number-changing processes are described by the last line in Eq.~\eqref{eq:hamiltonian}. We can therefore omit inelastic collisions by omitting the final line of this expression.
As we noted in Sec.~\ref{sec:Intro}, the evolution equation \eqref{eq:Schrodinger-like} is not a Schr\"odinger equation, both because of the absence of a prefactor $i$ on its right-hand side and because the ``state'' $w$ that $H$ acts on is also found inside $H$.  In a by-now-standard abuse of terminology, we nevertheless refer to $H$ as the Hamiltonian, and call its eigenvalues energies.

We will also choose the time-dependent scalings $A(y),D(y),$ and $r(y)$ in the same way as in Section 4 of~\cite{Rajagopal:2024lou}. That is, we choose $A(y)$ as suggested by number conservation,
\begin{equation}
\frac{\partial_y A}{A}=-3\frac{\partial_y D}{D}-1\,,
\end{equation}
although in our case number will not be exactly conserved. 
We choose $D(y)$ such that it decays towards the effective temperature $T_{\rm eff} = I_a/I_b$, 
\begin{equation}
\frac{\partial_y D}{D}=10\left( 1-D \left< \frac{2}{p} \right>\right)\,.
\end{equation}
%{\color{blue}where $\braket{\cdot}$ means $(\int_{\bm{p}} \cdot\; f)/(\int_{\bm{p}} f)$.}
Finally, we choose $r(y)$ to try to maximize the extent to which $f$ is well-described by $\psi^{(R)}_{10}$. Specifically, we choose $\partial_y r$ at each time such that the $u^2$ moment of the Boltzmann equation \eqref{eq:boltzmanneqn} for $\psi_{10}^{(R)}$ matches what it would be if we had $w=\psi_{10}^{(R)}$, that is,
\begin{equation}
\int_{\bm{p}} u^2 (\partial_y +H)\psi_{10}^{(R)}=0\,,
\end{equation}
which for $H$ given by Eq.~\eqref{eq:hamiltonian} gives us
\begin{equation}
    \partial_y r=\frac{1}{r}\frac{J_0}{J_4J_0-J_2^2}\left(
 2(J_2 - J_4) - 
   \frac{\tau \lambda_0 \ell_{\rm Cb} I_a}{D^2}(J_0 - 3 J_2)\right)
\end{equation}
regardless of whether or not inelastic collisions are included in $H$, where we have defined $J_n = \int_{-1}^1 du\; u^n e^{-u^2r^2/2}$. The first term in parentheses drives $r$ to larger values (larger anisotropy) and is dominant at early times, while the second term drives $r$ to smaller values and grows in importance over time, eventually causing isotropization of our basis.

We will use the same initial condition as in  Section 4 of Ref.~\cite{Rajagopal:2024lou}, namely
\begin{equation} \label{eq:init-cond}
    f({\bf p}, \tau = \tau_I) = \frac{\sigma_0}{g_s^2} e^{- \sqrt{2} p/Q_s} e^{- r_I^2 u^2 /2 } Q_0^{(R)}(u;r_I)
\end{equation}
with $\tau_I Q_s = 1$. We shall specify our choices for the parameters $\sigma_0$ and $r_I$ below.

As described in Ref.~\cite{Mazeliauskas:2018yef}, for moments defined by 
\begin{equation}\label{eq:moments-defn}
    \mathcal{N}_{b,c}(\tau)= \int_{\bm{p}} p_\perp^b |p_z|^c f(\bm{p})\,,
\end{equation}
the scaling exponents obtained from solving
\begin{equation}\label{eq:momentscalingequation}
    \partial_y \log \mathcal{N}_{b,c}= \alpha(\tau) - (b+2)\beta(\tau) -(c+1)\gamma(\tau)
\end{equation}
for any set of three independent moments $(b,c)$ will yield the same $\alpha$, $\beta$, and $\gamma$ if $f$ is scaling. By extracting the scaling exponents this way, we can test whether $f$ is approximately scaling. In all figures depicting scaling exponents below, we choose the following three sets of three moments: 
$(b,c)\in$\{(1,2), (0,2), (1,0)\} or \{(0,0), (0,2), (2,0)\} or  \{(2,2), (1,2), (2,0)\}. 

Finally, to substantiate our claims of observing both pre-hydrodynamic attractor behavior and hydrodynamization in the more strongly coupled examples considered in Sec.~\ref{sec:strongcoupling}, we will calculate the longitudinal pressure $P_L$ in our kinetic theory for systems initialized at various initial anisotropies (as parametrized by the initial value of the scale $r$). We compare this to the prediction of 1st order hydrodynamics by matching the energy density $\varepsilon$ from kinetic theory at the final time and integrating
\begin{align}\label{eq:1storderhydro}
    \partial_y \varepsilon = -\frac{4}{3}\varepsilon + \Phi
\end{align}
backwards, taking $\Phi=\frac{4\eta}{3\tau}$. The longitudinal pressure is $P_L=\frac{1}{3}\varepsilon -\Phi$. The shear viscosity is calculated as $\eta = T^3 \tilde{\eta}$ where we use $T=I_a/I_b$ (the effective temperature) and for each choice of coupling we choose $\tilde{\eta}$ such that $\partial_y \varepsilon$ matches the slope numerically obtained from kinetic theory at the final time.

\subsection{Prehydrodynamic Attractors at Very Weak Coupling}\label{sec:weakcoupling}

\begin{figure}
\centering
\includegraphics[width=\textwidth]{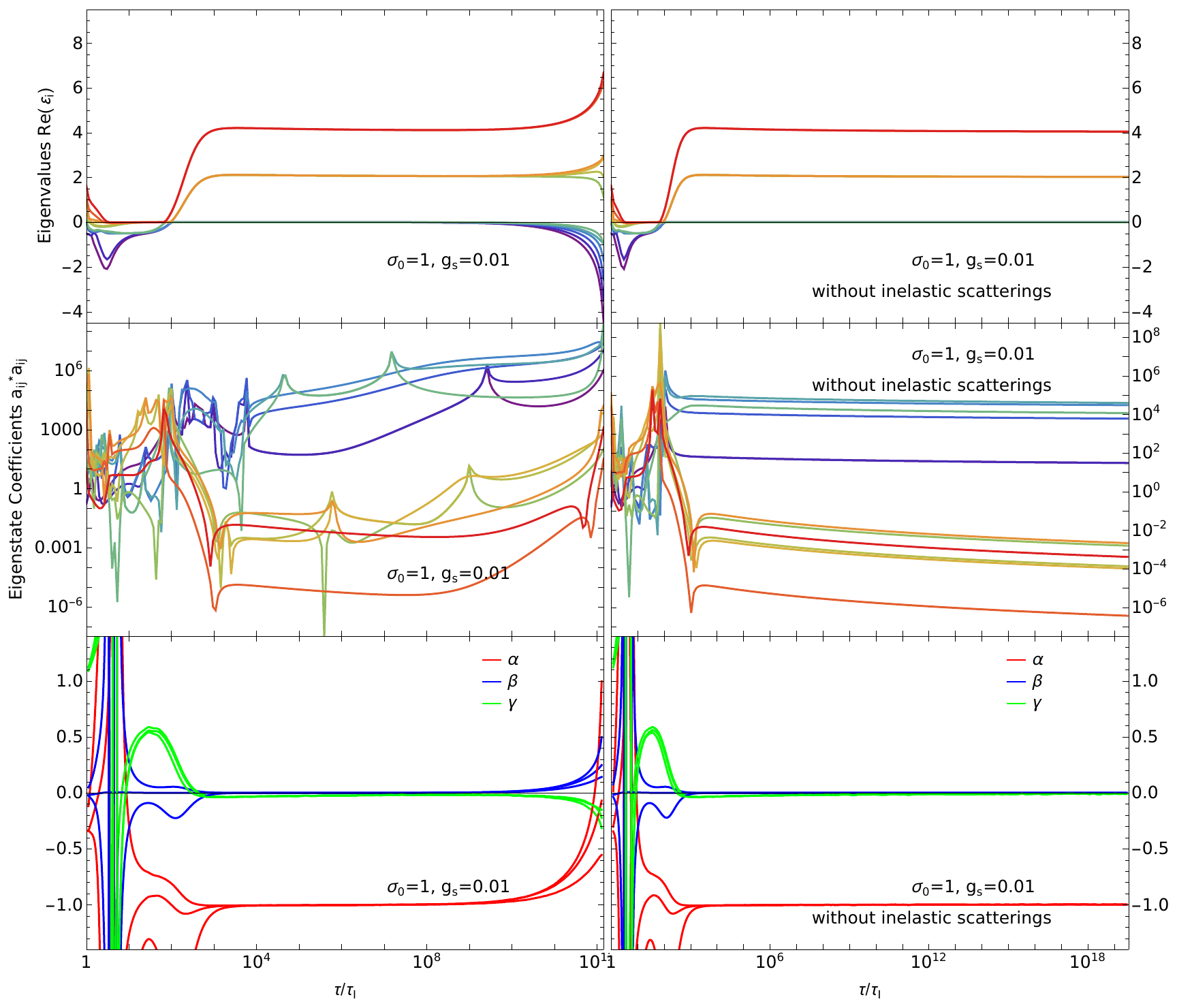}
\caption{
Effective energies Re[$\varepsilon_i$], coefficients $a_i$ of the eigenstates of $H$, and scaling exponents (as defined in Eq.~\eqref{eq:momentscalingequation} and plotted in each case for three different choices of $(b,c)$ in Eqs.~\eqref{eq:moments-defn} and \eqref{eq:momentscalingequation}) for the initial condition in Eq.~\eqref{eq:init-cond} with $g_s=10^{-2}$. On the left we use the full effective Hamiltonian $H$ as it appears in Eq.~\eqref{eq:hamiltonian} to evolve the rescaled distribution function $w$, while on the right we omit the terms corresponding to inelastic scattering. With or without inelastic processes, the system approaches the pre-thermal ``dilute" scaling state with its corresponding energy levels on approximately the same time scale $\tau/\tau_I \sim 10^3$. The eigenstate coefficients corresponding to the excited state bands quickly die off relative to the ground state band as a gap opens and the system ``falls'' onto the dilute attractor surface. Without inelastic scattering, the system dilutes forever and there is no sign that it ever hydrodynamizes. When inelastic scatterings are included, our basis is such that at late times we are unable to follow the dynamics of the system. This loss of  control of the calculation happens prior to hydrodynamization. %{\color{magenta} BSH: what is the number of basis states? (we can also state this in the text)}
}\label{fig:gs0.01}
\end{figure}

\begin{figure}
\centering
\includegraphics[width=\textwidth]{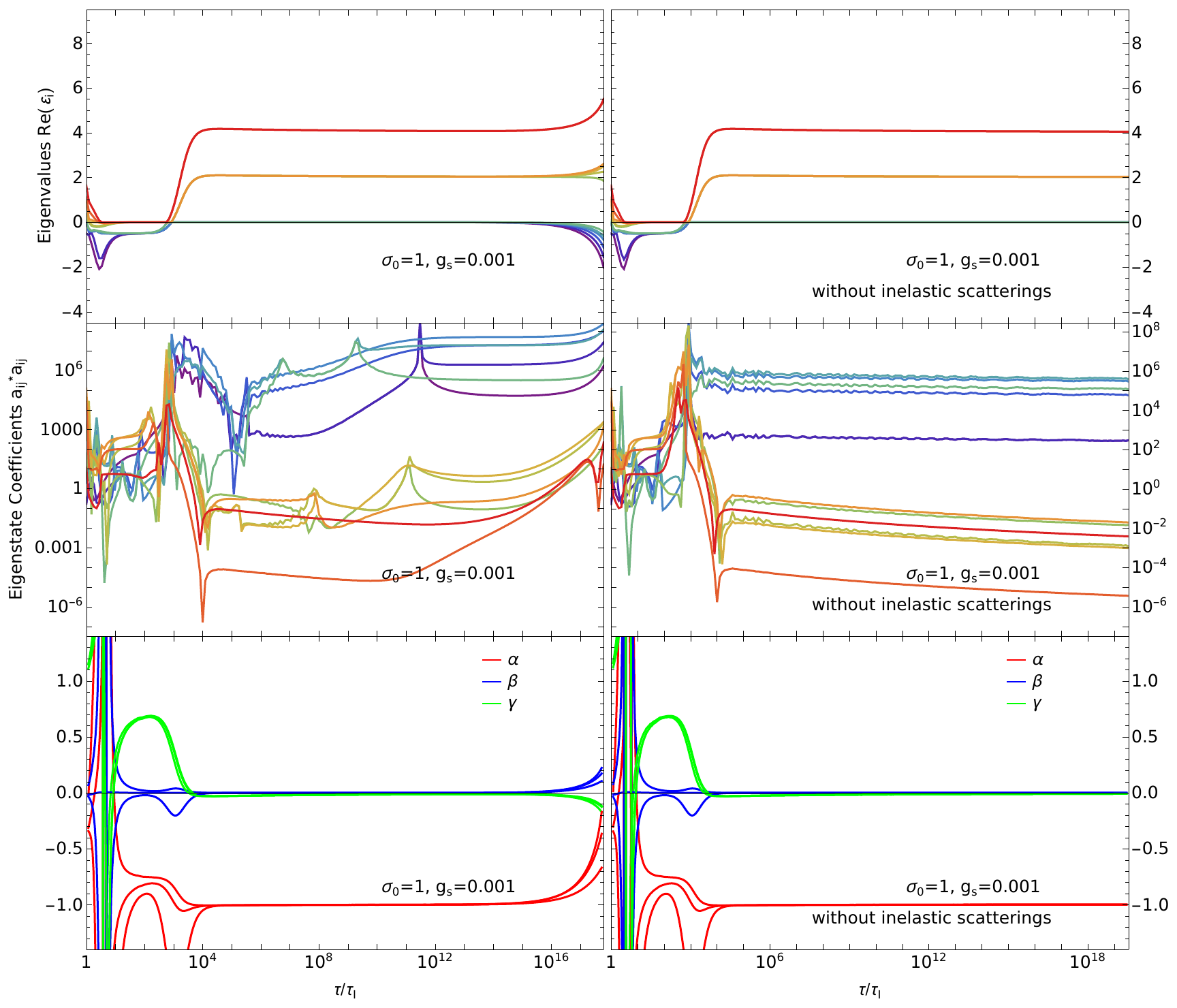}
\caption{As in Fig.~\ref{fig:gs0.01}, effective energies Re[$\varepsilon_i$], coefficients $a_i$ of the eigenstates of $H$, and scaling exponents for the initial condition in Eq.~\eqref{eq:init-cond} --- here with $g_s=10^{-3}$. Inelastic scatterings are omitted for the results in the panels on the right. The behavior is strikingly similar to Fig.~\ref{fig:gs0.01}, but for the difference in the timescales on which the system approaches the prethermal attractor and on which the calculation loses control when inelastic processes are included.}\label{fig:gs0.001}
\end{figure}

Results for the evolution of the kinetic theory including (left panels) and without (right panels) inelastic processes for the very weak coupling choices $g_s=10^{-2}$ and $10^{-3}$ with $r_I=\sqrt{3}$ and $\sigma_0=1$ are shown in Figs.~\ref{fig:gs0.01} 
and~\ref{fig:gs0.001}, respectively. 
With these very weak values of the coupling, the evolution of the scaling exponents, eigenvalues of $H$, and eigenstate coefficients resemble each other strikingly closely in shape with the value of the coupling seeming only to affect the timescale over which they evolve.
In each of these cases we observe that the ``dilute" fixed point ($\alpha=1,\beta=\gamma=0$) describes the system state for most of the solvable evolution, with effective energy levels closely matching the analytical prediction of Ref.~\cite{Brewer:2022vkq}. We neither expect nor find the BMSS fixed point ($\alpha=-2/3,\beta=0,\gamma=1/3$) in this case, due to the omission of the Bose enhancement factor from $I_a$. (We have already seen in Ref.~\cite{Rajagopal:2024lou} that the BMSS fixed point at early times is described naturally in the AH framework with Bose factors included, at least at weak coupling, upon assuming that inelastic processes are not yet important.) Note that the timescale at which the scaling exponents transition from being approximately free-streaming to the expected pre-thermal attractor values is $\tau/\tau_I \sim Q_s \tau \sim 1/g_s$, the timescale at which collisions begin to compete with expansion in the effective Hamiltonian.

\begin{figure}
    \centering
    \begin{subfigure}{0.8\textwidth}
        \centering
        \includegraphics[width=\textwidth]{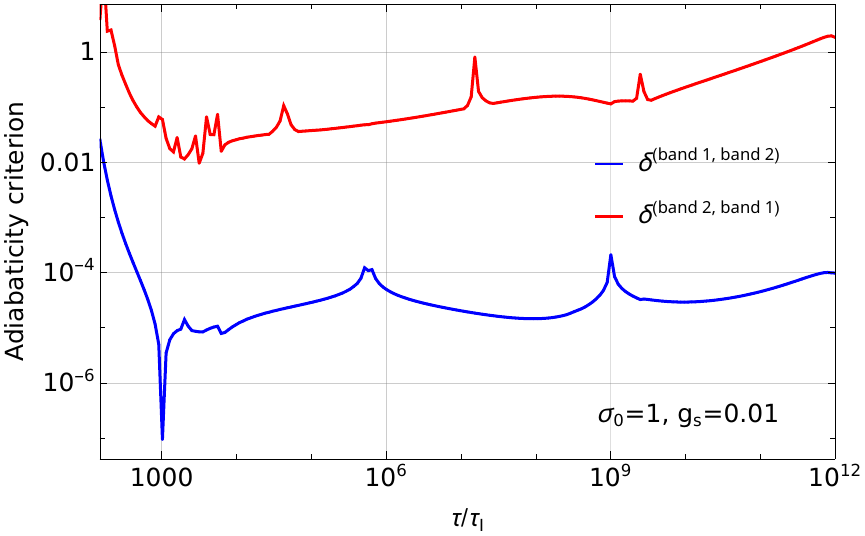}
    \end{subfigure}%
    \hfill
    \begin{subfigure}{0.8\textwidth}
        \centering
        \includegraphics[width=\textwidth]{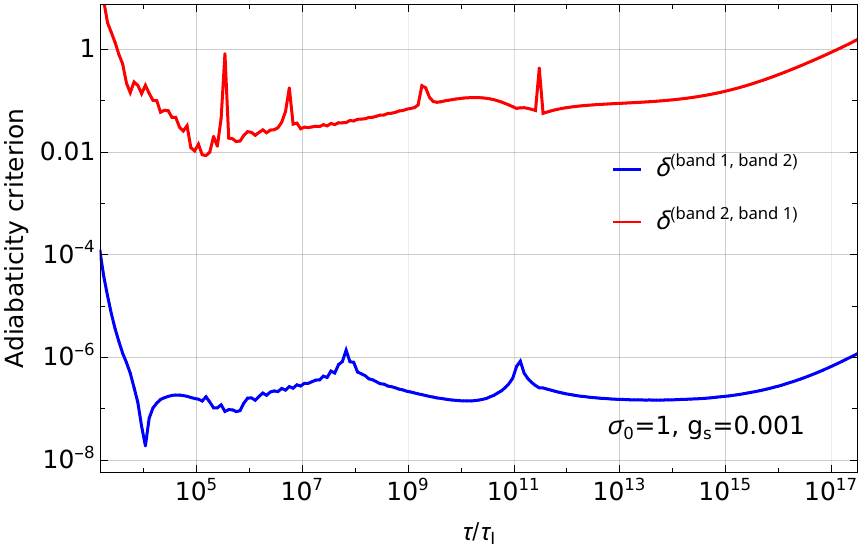}
    \end{subfigure}
    \caption{Band adiabaticity condition as defined in Eq.~\eqref{eq:adiabaticitycondition} for $g_s=10^{-2}$ (upper panel) and $10^{-3}$ (lower panel) during the evolution of the kinetic theory including both elastic and inelastic scattering, 
    plotted during the broad range of time in which the effective energy levels of $H$ are grouped into bands associated with dilute scaling. 
    In blue is $\delta_{\rm max}^{(\rm band~1,band~2)}$, while in red is $\delta_{\rm max}^{(\rm band~2,band~1)}$ where ``band 1" is the ground state band, while ``band 2" is the first excited state band. 
    %{\color{red}{The modes with the highest two energies are excluded from either band.}} 
    The smallness of the blue criterion quantifies the degree to which the ground state band evolves adiabatically, without influence from or mixing with the excited state band. The smallness of the red criterion quantifies the independence of the excited state band from the ground state band. Also, the smaller the red criterion the more rapidly the occupation of the states in the excited state band will die away, meaning the more rapidly the system is driven into some superposition of states from the ground state band. Either criterion for either coupling is consistently well below 1 during the dilute regime, and therefore the adiabatic explanation for pre-thermal attractor behavior is justified.}\label{fig:adiabaticity-weak}
\end{figure}

As observed in Ref.~\cite{Rajagopal:2024lou}, although 
during the ``dilute" stage of evolution the effective Hamiltonian $H$ does not feature a single lowest eigenvalue separated from all others by a gap, the evolution of the system nevertheless admits an adiabatic interpretation. 
Rather than a non-degenerate ground state, we see a set of states with approximately zero eigenvalues of $H$, namely  a ground state ``band", while the higher-energy states with a substantial gap will tend to decay away. The result is an attractor surface spanned by these degenerate ground state modes. We substantiate this claim by calculating the maximum adiabaticity criterion between energy bands,
\begin{equation}\label{eq:adiabaticitycondition}
    \delta^{(\text{band } i,\text{band } j)}_{\rm max} \equiv \max_{\substack{n \in \text{band i}\\m \in \text{band j}}} \left|\frac{\braket{n|_L \partial_y|m}_R}{\braket{m|_L |m}_R(\varepsilon_n-\varepsilon_m)}\right|\ ,
\end{equation}
where $\ket{m}_R,\bra{n}_L$ are respectively right and left instantaneous eigenstates of $H$, which we can find by diagonalizing $H$ on our basis at each time step, and $\varepsilon_{n/m}$ are the corresponding effective energy levels. (In practice, when we evaluate \eqref{eq:adiabaticitycondition} we exclude the modes with the highest two eigenvalues from either band.) As seen in Fig.~\ref{fig:adiabaticity-weak}, the band adiabaticity criterion is indeed small --- especially so for the effects of excited states on the ground state band --- thus confirming that the latter evolves adiabatically.

It is apparent that for the case of very weak coupling, adding inelastic scatterings has very little effect until very late times, at which point 
in the absence of inelastic processes the system continues to dilute uneventfully whereas in their presence
we seem to lose control of our description of the dynamics.
This is likely a limitation of the basis truncation we have used. That is to say, the dynamics is driving $f$ into a regime for which our basis is not well-adapted. In contrast, the solution without inelastic scatterings never seems to hydrodynamize (nor to break down). We can speculate that the breakdown  of our truncated calculation may in some way be related to the onset of hydrodynamization, which, at weak coupling, begins very far from a thermal distribution due to the strong dilution of the previous stage. 
Our truncated basis may not be able to capture the large changes associated with hydrodynamization from such a dilute and anisotropic state. Furthermore, in the transition between the pre-thermal ground state and the hydrodynamizing ground state, the ground state is \textit{necessarily} not scaling. Consequently, it makes sense that a truncated basis description of the distribution function may struggle to maintain control of $f$ during the transition between ground state forms of the dilute and thermal regimes. 
These challenges do not arise in the absence of inelastic processes, since in that case the kinetic theory shows no signs of hydrodynamization.
We also note that 
adding $1 \leftrightarrow 2$ processes should 
accelerate hydrodynamization more at stronger coupling, meaning that at stronger coupling the system should hydrodynamize before it dilutes as much as it does here.  We shall return to this point in the next Subsection.

\subsection{Hydrodynamization}\label{sec:strongcoupling}

\begin{figure}
\centering
\vspace{-0.4in}
\includegraphics[width=\textwidth]{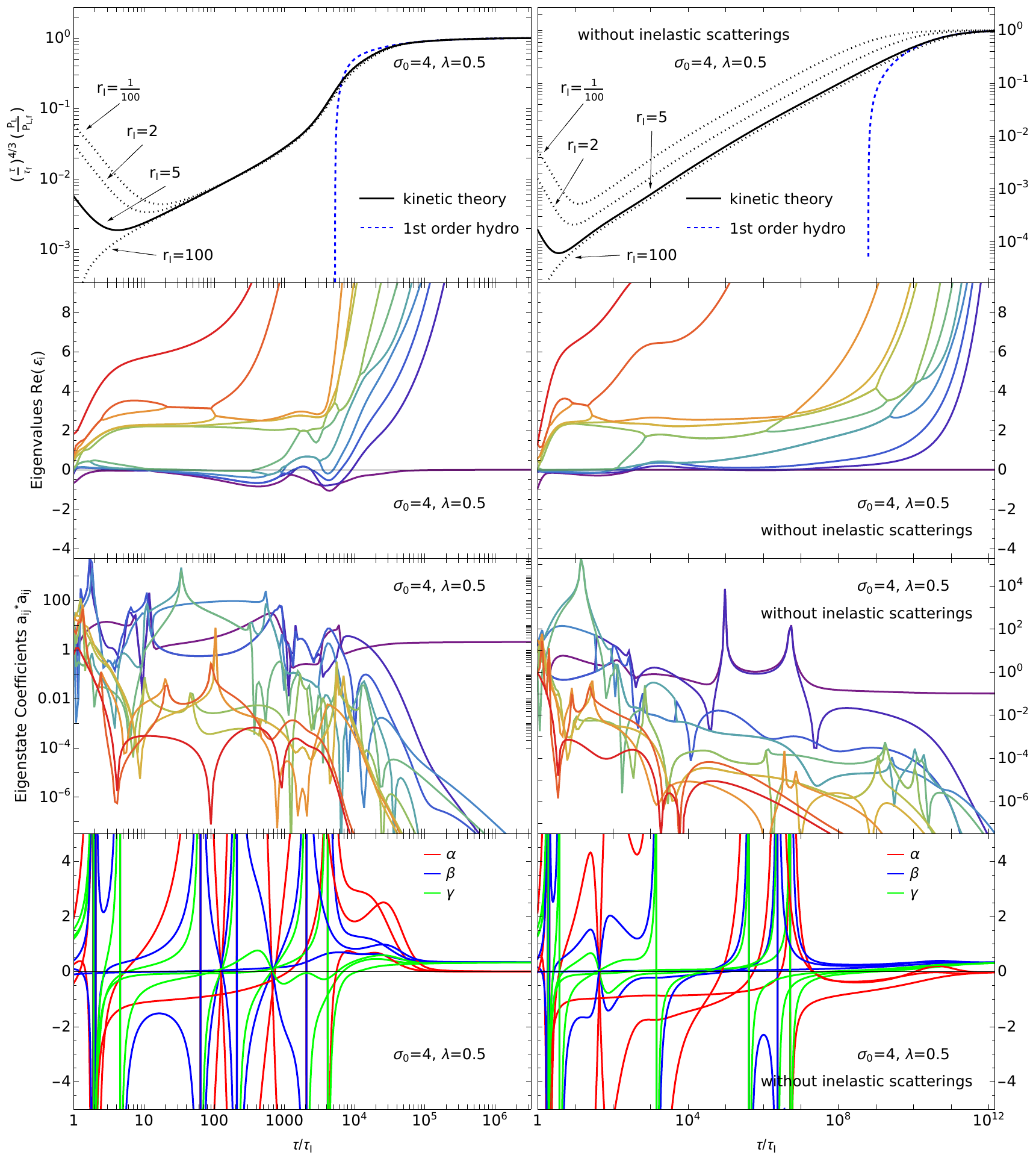}
\caption{Longitudinal pressure relative to its late-time hydrodynamic $\tau^{4/3}$ scaling (top panels, first order hydro as described around Eq.~\eqref{eq:1storderhydro}), effective energies Re[$\varepsilon_i$], $\varepsilon_i$ being the eigenvalues of $H$ (second row), coefficients $a_i$ of the eigenstates of $H$ (third row), and putative scaling exponents defined from Eqs.~\eqref{eq:moments-defn} and \eqref{eq:momentscalingequation} (fourth row) for the initial condition in Eq.~\eqref{eq:init-cond} in the kinetic theory with $\lambda=0.5$ and $\sigma_0=4$. Inelastic processes are omitted in the panels on the right. Effective energies and corresponding eigenstate coefficients are color-coded to match. In the plot of longitudinal pressure, curves for differing initial anisotropies are given to demonstrate attractor behavior. Inspecting the effective energies Re[$\varepsilon_i$], with or without inelastic scatterings, there are bands of states clustered around 0 and 2, which is the expected band gap for the dilute scaling regime. And, we see (third row) that as the system enters this regime the occupation of states in the excited band drops and (top row) the longitudinal pressure falls onto a common pre-hydrodynamic curve (top-left) or curve with a common shape (top-right).
However, 
we see (fourth row) that 
during this regime the ``scaling exponents'' do not scale at all, with
scaling setting in only much later, after hydrodynamization.}\label{fig:lambda0.5}
\end{figure}

\begin{figure}
\centering
\includegraphics[width=\textwidth]{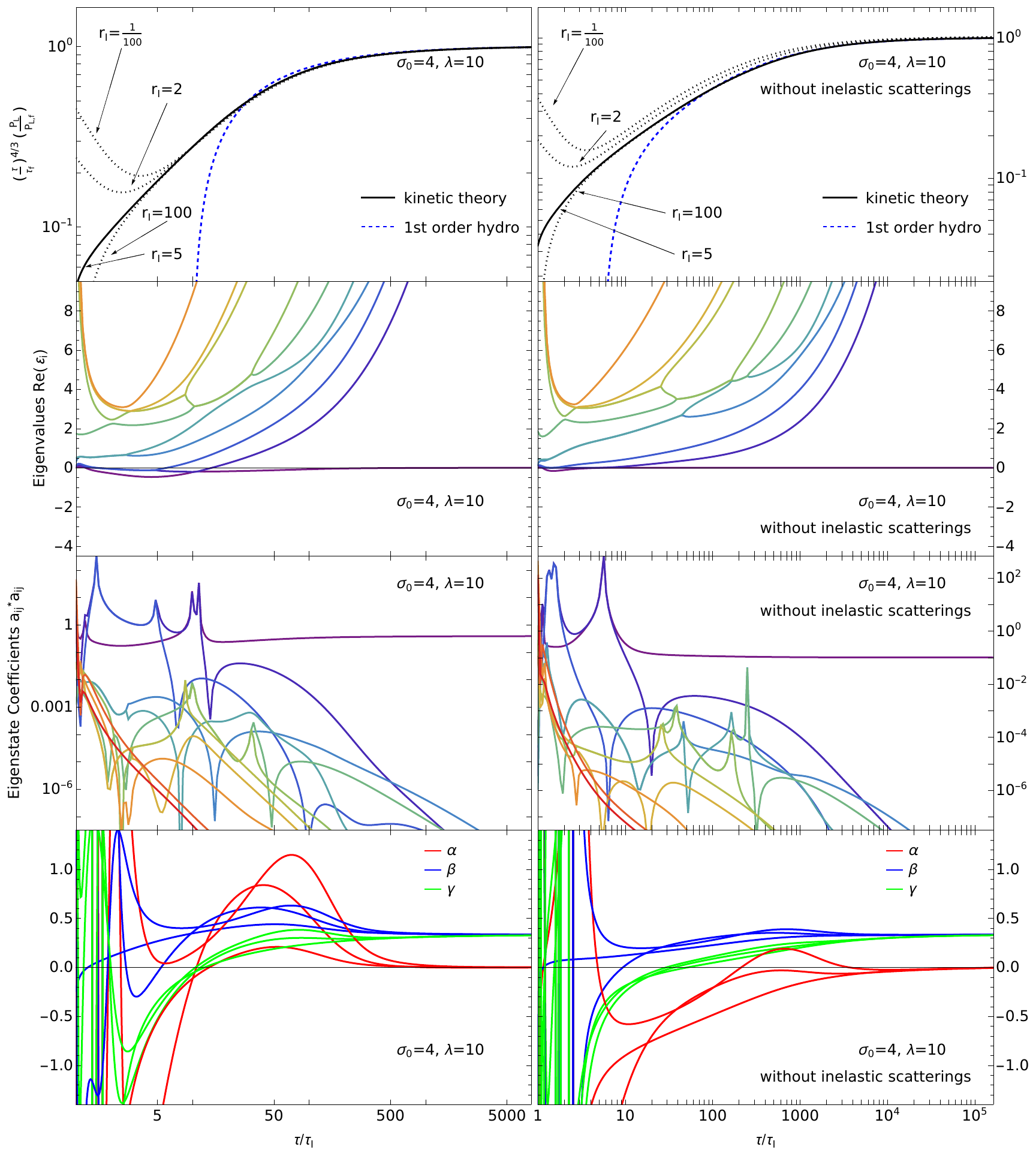}
\caption{As in Fig.~\ref{fig:lambda0.5}, longitudinal pressure, effective energies Re[$\varepsilon_i$], coefficients $a_i$ of the eigenstates of $H$, and putative scaling exponents for the initial condition in Eq.~\eqref{eq:init-cond} with $\sigma_0=4$ as before but now with the stronger, approximately realistic, coupling $\lambda=10$, corresponding to $\alpha_s\approx 0.27.$ Inelastic scatterings are omitted for the results in the panels on the right. As compared to Fig.~\ref{fig:lambda0.5}, it is much harder to argue that the ``dilute" scaling regime is present 
%either 
by inspection of 
%the scaling exponents or 
the effective energy levels, and the addition of inelastic scatterings has no noticeable effect on the timescale of hydrodynamization. Nonetheless, there is still some attractor behavior in the sense that differing initial conditions converge prior to the applicability of 1st order hydrodynamics, and this can be associated with the faster decay of the highest excited state modes. As in Fig.~\ref{fig:lambda0.5}, the putative scaling exponents only show scaling behavior well after hydrodynamization.}\label{fig:lambda10}
\end{figure}

Results for the evolution of the kinetic theory are shown in Figs.~\ref{fig:lambda0.5} and \ref{fig:lambda10} for $\lambda=0.5$ and $\lambda=10$ respectively, where $\lambda$ is the t'Hooft coupling $\lambda\equiv N_c g_s^2$, and with $r_I=5$ and $\sigma_0=4$ for both couplings. These correspond to $g_s=0.41$ and $g_s=1.83$, respectively. 
As anticipated, including inelastic scatterings substantially quickens the onset of hydrodynamization.  
For example, with $\lambda=0.5$ a gap opens up above a unique lowest eigenvalue mode in the spectrum of eigenvalues of $H$ and the longitudinal pressure is described well by first-order hydrodynamics for $\tau > \tau_{\rm hyd}\sim 10^{9-10} \tau_I$ in the absence of inelastic processes
and for $\tau > \tau_{\rm hyd}\sim 10^4 \tau_I$ when we incorporate inelastic processes in the kinetic theory. 
The onset of hydrodynamization 
reaches physically reasonable timescales (i.e., hydrodynamization times $\tau_{\rm hyd} \sim (10-20)\tau_I=(10-20)/Q_s$) in the case with a physically reasonable coupling $\lambda=10$.  This confirms that, as first demonstrated in Ref.~\cite{Kurkela:2015qoa}, QCD kinetic theory -- which of course includes inelastic processes -- can describe hydrodynamization on a realistic timescale when the coupling is increased to realistic values.

With or without inelastic scattering, and for either coupling, it is satisfying to see in Figs.~\ref{fig:lambda0.5} and \ref{fig:lambda10} that we have a clear hierarchy in the occupancy of modes well before hydrodynamization. 
At early times a small set of modes with low-lying eigenvalues of $H$ dominate, and then at later times (not much later in the case where $\lambda=10$)
a gap opens above a single lowest eigenvalue mode, with the opening up of this gap
driving the system to the hydrodynamizing mode -- which is the unique late-time ground state. Furthermore, the time at which a clear gap opens (see second row of panels in Figs.~\ref{fig:lambda0.5} and \ref{fig:lambda10}) roughly coincides with the apparent time of hydrodynamization (see first row of panels).

\begin{figure}
    \centering
    \includegraphics[width=0.8\linewidth]{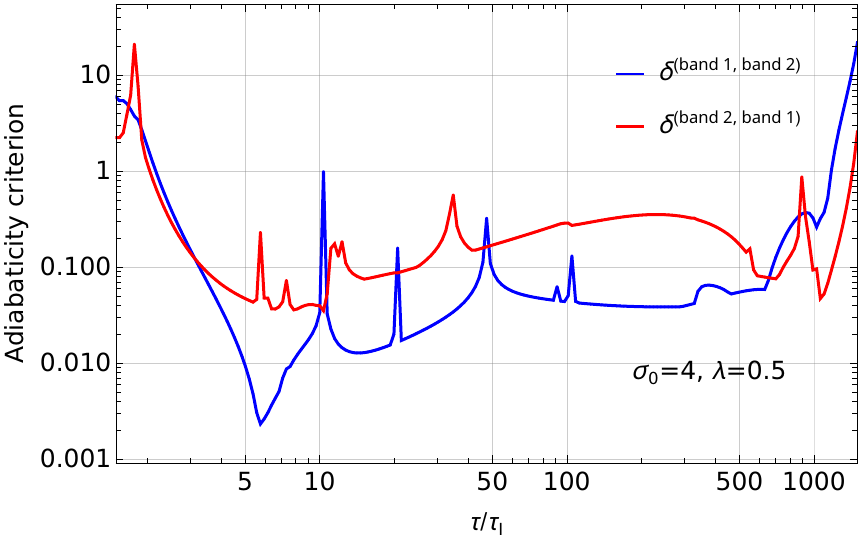}
    \caption{Band adiabaticity conditions as defined in Eq.~\eqref{eq:adiabaticitycondition} for the evolution of the kinetic theory including both elastic and inelastic scattering processes with $\lambda=0.5$ and $\sigma_0=4$  during the time window in which the effective energy levels of $H$ resemble the bands associated with dilute scaling, with the colors and bands defined as in Fig.~\ref{fig:adiabaticity-weak}. Both band adiabaticity conditions are indeed well below 1, and are particularly small during the early part of the dilute regime when most of the decay of excited states occurs. 
    }
    \label{fig:strongadiabaticity}
\end{figure}

It is particularly interesting to note that even though $f$ is clearly {\it not} scaling for these stronger couplings -- note the apparent chaos in the bottom rows of both
Figs.~\ref{fig:lambda0.5} and \ref{fig:lambda10} prior to very late times well after hydrodynamization -- 
the intuition coming from the use of the AH framework remains relevant. This is most clearly seen in Fig.~\ref{fig:lambda0.5}, where 
 there are clear
energy bands (second row of panels) resembling those expected for the dilute scaling regime over a wide range of times from before $10\tau_I$ to after $1000\tau_I$ --- well before hydrodynamization --- even though the ``scaling exponents'' are not scaling at all during this epoch (bottom row of panels).  The spectrum of eigenvalues of $H$ is telling us, loud and clear, that there is a pre-hydrodynamic attractor --- no matter the lack of pre-hydrodynamic scaling.
We can confirm this via observing the decay of the corresponding excited state coefficients in the third row of panels. We further confirm the validity of the AH picture in Fig.~\ref{fig:strongadiabaticity}, where we see that the evolution of $H$ is reasonably adiabatic in this time window as defined by our band adiabaticity condition \eqref{eq:adiabaticitycondition}. This suggests that $f$ during this time window is dominated by a small number of low-effective-energy modes, but these modes do not take the scaling form of Eq.~\eqref{eq:scalingform}.
That is to say, even though pre-thermal attractor behavior in weakly coupled theories has previously typically been understood through the lens of pre-thermal scaling, this example demonstrates that scaling is not a prerequisite for attractor behavior and that adiabatic hydrodynamization is more general: it can describe the pre-hydrodynamic loss of memory that is the defining characteristic of attractor behavior even in the absence of scaling.

We can also see that if and only if inelastic processes are included at $\lambda=0.5$ and $\lambda=10$, it appears that the longitudinal pressure falls onto a universal curve in the pre-thermal attractor regime.  
In the kinetic theory with 
only elastic scatterings, the evolution of the longitudinal pressure for varied initial conditions shares a shape but does not fall onto a unique curve without a horizontal shift. (One can see from Ref.~\cite{Rajagopal:2024lou} that these curves do coincide at late times if each is suitably shifted in $\tau$.)
This highlights the fact that an attractor surface such as we see in the pre-thermal regime may or may not correspond to universal physical dynamics, 
depending on which observables are considered.
The collapse of varied initial states onto an attractor surface corresponds to the loss of much memory about the initial state, but not all.  Different initial states can end up in different linear superpositions of the eigenstates of $H$ that make up the pre-hydrodynamic attractor surface.
Only later when a gap within the pre-hydrodynamic band of ground states opens up, separating all but one from a unique ground state, does the system
hydrodynamize and completely lose all memory of its initial state other than via conserved quantities.
The fact that in the left-hand columns of Figs.~\ref{fig:lambda0.5} and \ref{fig:lambda10} (not very weak coupling; inelastic scatterings included) the curves {\it do} collapse onto one another during the pre-hydrodynamic epoch reflects the fact that the inelastic processes break the conservation of $n\tau$. 
As discussed above, the inclusion of the $1\leftrightarrow 2$ kernel also means that
the 
overall normalization $f(\bm{p}=0)$ of the distribution function is determined at late times
by $H$, not by the initial conditions. 
Although the inclusion of inelastic processes is a necessary condition for the longitudinal pressure evolution to be universal when the system is on the pre-hydrodynamic attractor surface, it is not sufficient: for example, this is not so at the weaker couplings considered in Sec.~\ref{sec:weakcoupling}.

\begin{figure}
    \centering
    \includegraphics[width=0.7\linewidth]{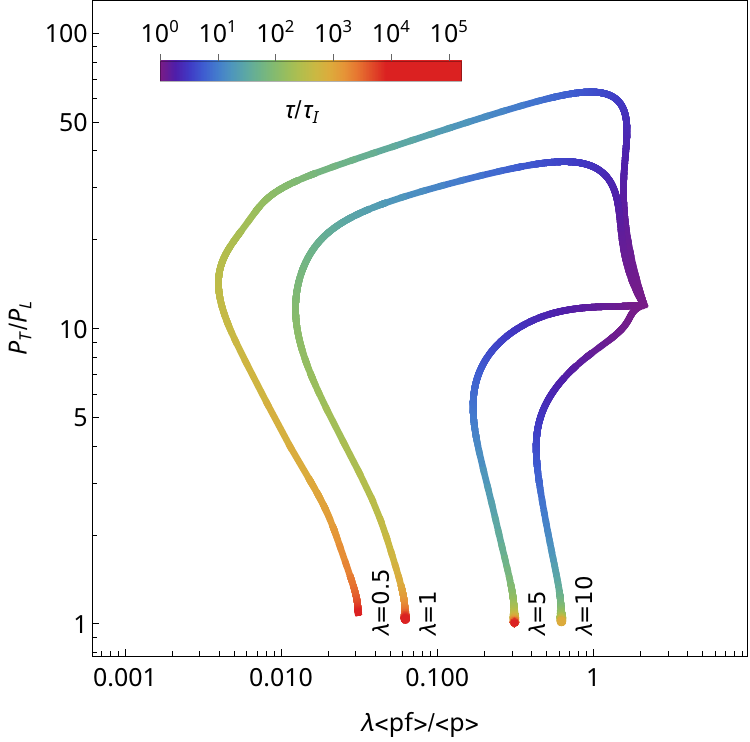}
    \caption{Pressure anisotropy and occupancy as a function of time for $\sigma_0=4$ during the hydrodynamization of four kinetic theories that include inelastic processes with differing choices of coupling. In all cases, the kinetic theory is initialized with the same anisotropy $P_T=12 \,P_L$ and the same occupancy. 
    The curves show how the anisotropy and occupancy change during hydrodynamization, with the stages of hydrodynamization most apparent in the two theories with smaller couplings. Once the system has fallen onto the dilute prehydrodynamic attractor surface, it dilutes and isotropizes somewhat -- the ``slanted top'' of the trajectories for $\lambda=0.5$ and $1$. Later, after the curves have turned to the right (occupancy increasing) and dropped rapidly (largely isotropized), the system is attracted to the hydrodynamizing attractor. As compared to the results plotted in the analogous figure in Ref.~\cite{Kurkela:2015qoa}, the systems here dilute slightly more before hydrodynamizing, which is likely due to the absence of the secondary hard splittings that play some role in bottom-up thermalization but are absent in our kinetic theory because of the soft approximation that we made in evaluating the 
    $1\leftrightarrow 2$ collision kernel \eqref{eq:1to2final}. Our results differ significantly from those in Ref.~\cite{Kurkela:2015qoa} only at very early times, before the kinetic theory reaches the pre-hydrodynamic attractor. This difference is explained well by the fact that we have omitted all Bose enhancement terms in our kinetic theory, which becomes unimportant once the occupancy drops below 1. 
    }
    \label{fig:ani-vs-occ}
\end{figure}

We close this Section by looking at what is happening in each phase of hydrodynamization by plotting the 
trajectory of the system through the plane of pressure anisotropy vs. energy-weighted occupancy, an approach that has yielded considerable insight since the work of Ref.~\cite{Kurkela:2015qoa}. This is shown in Fig.~\ref{fig:ani-vs-occ}. 
Let us first look at the evolution of the kinetic theory with $\lambda=0.5$, as in Fig.~\ref{fig:lambda0.5}.  We see in Fig.~\ref{fig:ani-vs-occ} that in this case the system initially rapidly anisotropizes as it falls onto the pre-hydrodynamic attractor (actually, although this cannot be seen in this representation, the pre-hydrodynamic attractor surface).
We see from Fig.~\ref{fig:ani-vs-occ} that while it is on the pre-hydrodynamic attractor (as indicated by the dilute band structure of the energy
levels in Fig.~\ref{fig:lambda0.5}) the system dilutes and slightly isotropizes, all prior to hydrodynamization. 
Later, the system rapidly isotropizes and becomes less dilute, and hydrodynamization begins.
The dynamics for $\lambda=1$ is similar, but when we look at $\lambda=10$ in Fig.~\ref{fig:ani-vs-occ}
we see no clear regime of dilution-while-slowly-isotropizing,
which is consistent with the lack of a distinct 
epoch with dilute band structure in the eigenvalues in Fig.~\ref{fig:lambda10}.
The general shape of the trajectories in Fig.~\ref{fig:ani-vs-occ} at various couplings --- and particularly the presence of a distinct phase of dilution for weaker couplings --- is in excellent agreement with the results first obtained in Ref.~\cite{Kurkela:2015qoa}. Here, we can directly connect it to the dilute pre-thermal attractor surface identified in Ref.~\cite{Brewer:2022vkq} and further analyzed in Ref.~\cite{Rajagopal:2024lou}.

We also note that for the {\it very} weak couplings of Figs.~\ref{fig:gs0.01} and \ref{fig:gs0.001}, if plotted in Fig.~\ref{fig:ani-vs-occ} the system would shoot up to very large anisotropies, and then follow the pre-hydrodynamic attractor surface far to the left (diluting enormously while isotropizing only modestly). As we have discussed above, hydrodynamization at such very weak couplings starts from a pre-hydrodynamic stage that is very strongly anisotropic, which presents challenges for our calculation using a truncated basis.  If these very weak couplings were phenomenologically relevant, it would be worthwhile to repeat the calculation with a different choice of basis designed for this circumstance.

The general qualitative and quantitative agreement of Fig.~\ref{fig:ani-vs-occ} with Ref.~\cite{Kurkela:2015qoa} also shows the physical reasonableness of the simplified kinetic theory considered in this work, particularly in contrast with the elastic-scattering-only theory considered in Ref.~\cite{Rajagopal:2024lou}. The only significant difference we see in comparing Fig.~\ref{fig:ani-vs-occ} to the analogous results in Ref.~\cite{Kurkela:2015qoa} is at early times when the occupancy is above 1. This is an expected difference resulting from our omission of Bose enhancement terms, and the expected scaling behavior and trajectory in occupancy and anisotropy can be easily restored by including specifically the Bose enhancement factor in $I_a$, as can be seen in Ref.~\cite{Rajagopal:2024lou}. Although we are still missing some of the qualitative features of bottom-up thermalization as described in Ref.~\cite{Baier:2000sb}, namely  secondary hard gluon splittings, it is reassuring to see that the physical picture of hydrodynamization (including a pre-hydrodynamic attractor surface and the hydrodynamizing attractor) is so fully described in such a unified fashion in the AH framework, as is hydrodynamization on a reasonable timescale with a reasonable (intermediate) value of the coupling. And, furthermore,
the sequential pre-hydrodynamic loss of memory that is the defining characteristic of hydrodynamization via an attractor surface followed by a unique attractor can arise with or without scaling, and is described equally naturally in the AH framework in either case.

\section{A Generic Feature of the Non-Hydrodynamic Sector of the Energy Spectrum at Late Times}\label{sec:hydro_inevitability}

With the results of the preceding Section, we have accomplished what we set out to do in terms of showing that the AH framework provides a unified description of hydrodynamization when inelastic processes are included, as it does in their absence, and with or without scaling: loss of memory takes place sequentially, and hydrodynamics is reached when one low-energy mode emerges, i.e.~when a unique ground state that subsequently evolves into a thermal momentum distribution takes over the dynamics.

\begin{figure}
    \centering
    \includegraphics[width=0.49\linewidth]{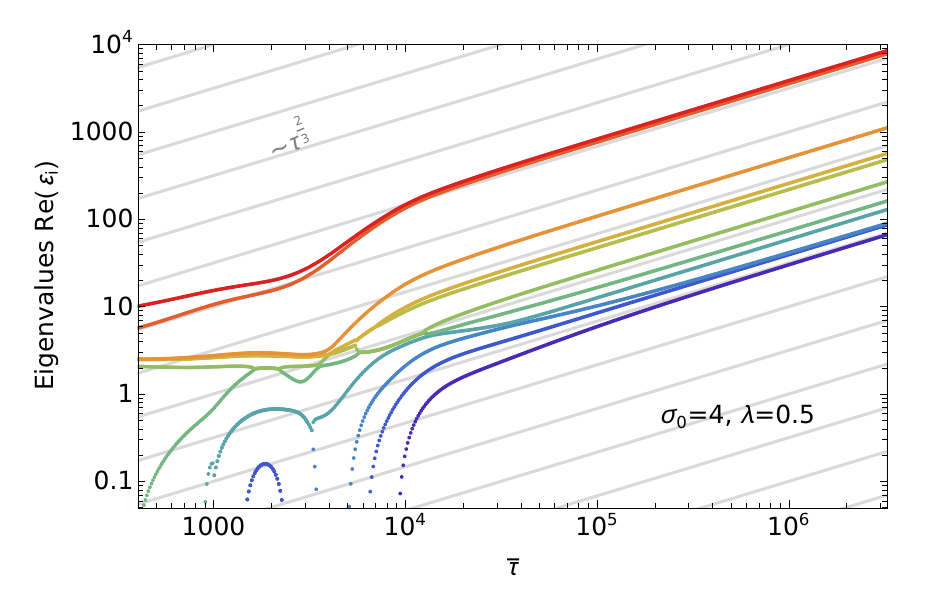}
    \includegraphics[width=0.49\linewidth]{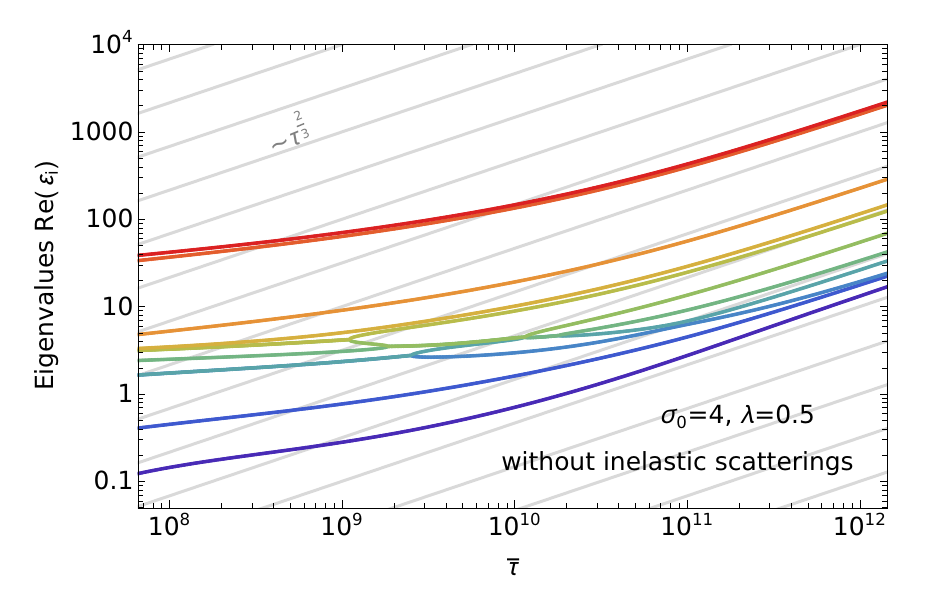}
    \includegraphics[width=0.49\linewidth]{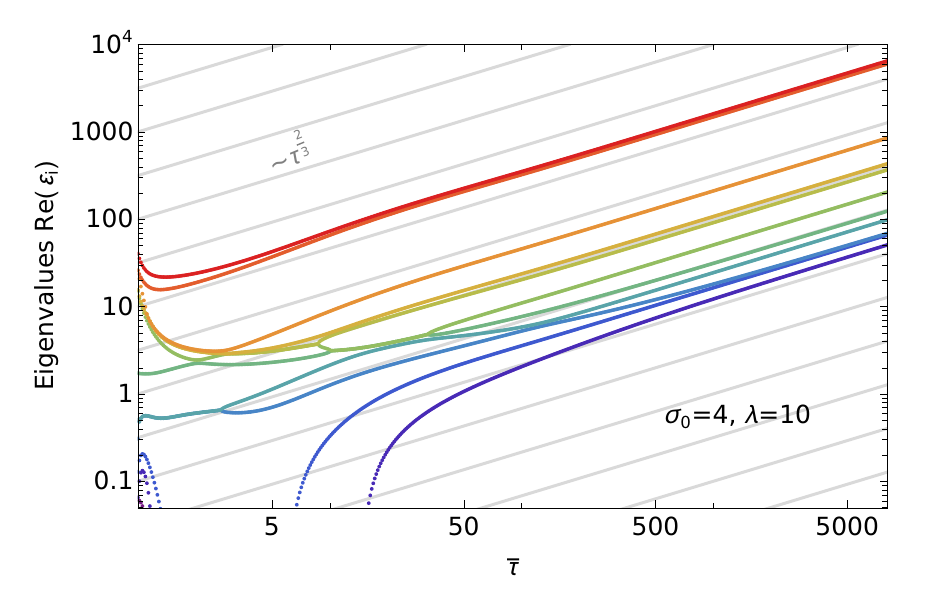}
    \includegraphics[width=0.49\linewidth]{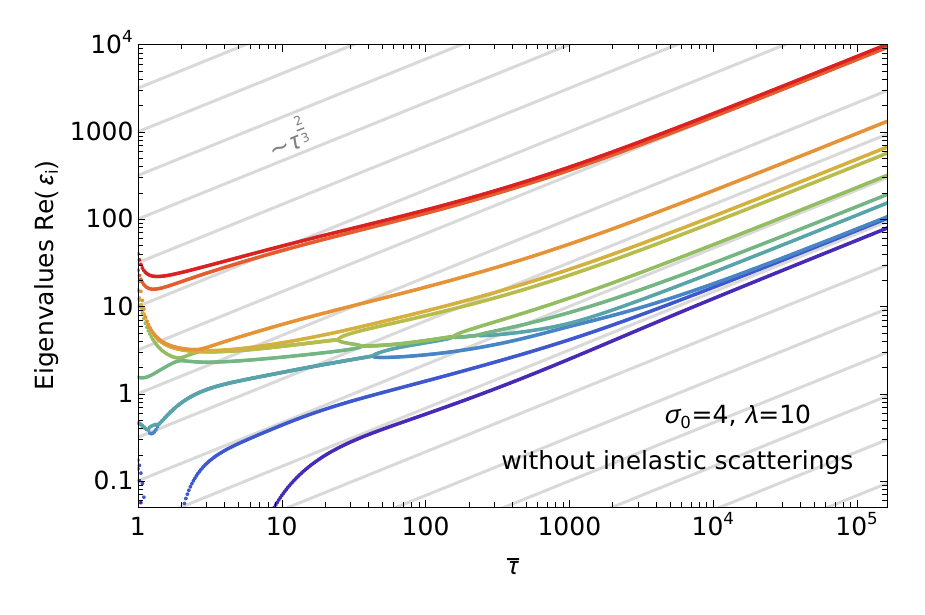}
    \caption{Energy levels previously shown in Figures~\ref{fig:lambda0.5} and~\ref{fig:lambda10}, other than the lowest eigenvalue, displayed on a logarithmic scale and compared with a slope of $2/3$.}
    \label{fig:elevels-Log}
\end{figure}

In this Section, we focus on an evident regularity of the late-time behavior of the eigenvalues of $H$ in Figs.~\ref{fig:lambda0.5} and \ref{fig:lambda10}.  In all four cases shown in these Figures ($\lambda=0.5$ and $10$; with and without inelastic processes) once hydrodynamization has begun, which is to say once a gap has opened up between the lowest and second-lowest eigenvalues of $H$,
it appears that this gap then grows exponentially in $y$, meaning as a power law in $\tau$. Furthermore, all of the eigenvalues of $H$ other than the lowest seem to grow in this fashion.
By replotting the eigenvalues on a logarithmic scale in Fig.~\ref{fig:elevels-Log}, we see that %\sout{at late times,}
once hydrodynamization has begun, all the eigenvalues of $H$ except for the lowest
grow rapidly, apparently exponentially, with $y$.
And, we see that at late times after the evolution is well described by hydrodynamics this growth
approaches $e^{2y/3} \propto \tau^{2/3}$.
We shall see below that this late time behavior is a consequence of ideal hydrodynamics with boost-invariant longitudinal expansion.
What we find interesting is that 
all of the non-hydrodynamic modes of the theory decouple in this way. 
The fact that this happens for both collision kernels (with and without inelastic processes) 
prompts us to ask whether this could be a general feature of the eigenvalue spectrum in any kinetic theory that has begun to hydrodynamize.
This is especially relevant given the implications that such behavior can have for the speed with which the system would re-equilibrate if it is perturbed or excited after it reaches the hydrodynamic regime---in other words, for how ``attractive'' the hydrodynamic attractor is.

We now show that this behavior holds for a broad class of kinetic equations describing boost-invariant longitudinal expansion,
\begin{equation}
    \partial_y f = p_z \partial_{p_z} f - \tau_I e^y C[f] \ , \label{eq:kin-sec4}
\end{equation}
provided that the following criteria hold:
\begin{enumerate}
    \item The kinetic theory describes massless particles, and therefore all of the dimensionful quantities in the collision integral are quantities derived from $f$. A formal 
    way of stating this is that $C[f]$ may be written as
    \begin{equation}
        C[f]({\bf p}) = \sum_{n=1}^\infty \frac{p^{1-3n}}{n!} \int_{{\bf p}_1 , \ldots , {\bf p}_n} C_n \!\left(\frac{{\bf p}_1}{p} , \ldots , \frac{{\bf p}_n}{p};\frac{{\bf p}}{p} \right) f({\bf p}_1) \cdots  f({\bf p}_n) \, , \label{eq:massless-scaling-C}
    \end{equation}
    where the functions $C_n$, which characterize the functional derivatives of $C[f]$ with respect to $f$, are dimensionless and only depend on dimensionless ratios of the momenta.
    All the kinetic theories that we have used in this paper have this form, as does QCD EKT with massless quarks.

    \item An isotropic solution $f_0$ to $C[f_0] = 0$ exists,  and the linearized spectrum of $C$ around this solution is bounded from below.  
    In practice, $f_0$ is the thermal equilibrium distribution of the theory (Boltzmann, Fermi-Dirac or Bose-Einstein). Given criterion 1, rescaling the momenta corresponds to rescaling the effective temperature that characterizes $f_0$.

    \item The isotropic solution $f_0$ is characterized by a single dimensionful scale $D(y)$ --- the effective temperature --- and
    once the distribution enters the regime where $f$ is close enough to $f_0$ that the collision kernel may be linearized around local thermal equilibrium, $D(y)$ 
    falls off more slowly than $1/\tau \propto e^{-y}$. We discuss this linearization in detail in what follows. (Note that $D(\tau)\propto 1/\tau^{1/3}$ in ideal hydrodynamics with boost-invariant longitudinal expansion, and viscous effects yield a $D$ that drops even more slowly than that. $D$ drops  more slowly than $1/\tau$ in any kinetic theory description of the approach to boost-invariant hydrodynamics that we know of.) 

\end{enumerate}
With these assumptions, if we 
choose $A(y)=1$ (see Eq.~\eqref{eq:A-w-definition})
we can write 
\begin{align}
    f(p,u,y) &=  
    f_0(p,y)+  
    \Delta f\left(p,u,y\right) \nonumber \\
    &=w_0\left(\chi\right) + \Delta w\left( \chi,u,y\right) \, , \label{eq:f-linearization}
\end{align}
where $\chi = p/D(y)$ as before. Upon assuming that $\Delta w$ is small compared to $w_0$, we may linearize the collision operator around $f_0 = w_0(p/D)$ and use criterion 2 to obtain 
\begin{align}\label{eq:linearizing-collision-operator}
    C[f]({\bf p}) &= C[f_0]({\bf p}) + \int_{{\bf k}} \left. \frac{\delta C[f]({\bf p})}{\delta f({\bf k})} \right|_{f = f_0} \Delta f({\bf k}) + \mathcal{O}(\Delta f^2) \nonumber \\
    &= 0 + D(y) L \, \Delta w + \mathcal{O}(\Delta w^2) \, ,
\end{align}
where $L$ is the time-independent linear operator that characterizes the collision kernel of the linearized theory. 
The second equality follows from the fact that 
\begin{equation}\label{eq:linearizing-collision-operator-2}
    \int_{{\bf k}} \left. \frac{\delta C[f]({\bf p})}{\delta f({\bf k})} \right|_{f = f_0} \Delta f({\bf k}) = D(y) \int_{{\bs \chi}} \left. \frac{\delta \tilde{C}[w]({\bs \chi})}{\delta w({\bs \chi})} \right|_{w = w_0} \Delta w({\bs \chi}) \equiv D(y) L \, \Delta w \, ,
\end{equation}
where we have used the scaling properties~\eqref{eq:massless-scaling-C} of the collision kernel in criterion 1 to rewrite $C[f]$ in terms of a rescaled collision kernel $\tilde{C}$ that takes $w$ as its argument, $C[f]({\bf p}) = D(y)\tilde{C}[w]({\bs \chi})$. We use the symbol $\tilde{C}$ to emphasize that it acts on the rescaled distribution $w$ rather than on $f$, even though the scaling symmetry in Eq.~\eqref{eq:massless-scaling-C} implies that the functional form of $C$ and $\tilde{C}$ is the same. This shows that $L$ does not depend on $D$.

We note that previous work has shown how this linear operator $L$ completely specifies the transport coefficients in the hydrodynamic regime of the theory~\cite{Jeon:1995zm,Arnold:2002zm,Arnold:2000dr,Arnold:2003zc}, and that the hydrodynamic equations of the theory with said transport coefficients can be derived from the linearized kinetic equation specified by $L$. (The canonical way of doing this would be the Chapman-Enskog method~\cite{chapman1990}. Other approaches to derive hydrodynamics have been developed recently in order to avoid known truncation issues~\cite{Denicol:2010xn,Martinez:2010sc,Florkowski:2010cf,Denicol:2012cn,Denicol:2016bjh,Bemfica:2019knx,Kovtun:2019hdm,Bemfica:2020xym,Bemfica:2020zjp}.)
In this way, the linearized theory contains hydrodynamics within it, but it also 
describes the non-hydrodynamic excitations of the complete kinetic theory in the near-equilibrium regime. Therefore, it can be used to study whether and how the dynamics of the two sectors becomes decoupled. 

After linearizing the collision operator as in \eqref{eq:linearizing-collision-operator}, the equation of motion for $\Delta w$ becomes
\begin{align}\label{eq:Delta-w-EoM}
    \partial_y \Delta w &= - H(y) \Delta w + K(y) w_0 \, , 
\end{align}
where
\begin{align}
    K(y) = \left( u^2 + \frac{\partial_y D}{D} \right) \chi \partial_\chi + u (1-u^2) \partial_u \, ,
\end{align}
and
\begin{equation}\label{eq:H-DK-L}
H(y) = \tau_I e^y D(y) L - K(y)\ . 
\end{equation}

As long as criterion 3 is satisfied ($D(y)$ drops more slowly than $\tau^{-1} = \tau_I^{-1} e^{-y}$), we conclude that at sufficiently late times 
(sufficiently large $e^y$), $\tau_I e^y D \gg 1$ meaning that the relative contribution of $K$ to the spectrum of $H$ is small and may be dropped and $H$ takes the simpler form
\begin{equation}\label{eq:late-time-H}
    H = \tau_I e^y D(y) L \ .
\end{equation}
This late-time simplification originates from the fact that the prefactor of  $D(y)L$  on the right-hand side of Eq.~\eqref{eq:H-DK-L} is greater than the prefactor of $K(y)$ by  $e^y$, which is a direct consequence of the fact that the expansion is boost-invariant.  
The late-time $H$
in Eq.~\eqref{eq:late-time-H} has time-independent eigenfunctions (which implies that their coefficients evolve adiabatically) determined by the eigenfunctions of the time-independent operature $L$. And,  the eigenvalues of this $H$ are given by $\tau_I e^y D(y) l$ where $l \geq 0$ is the corresponding eigenvalue of $L$.
Because of the scale symmetry of the collision kernel, 
$C[f_0+\Delta f]$ vanishes when $\Delta f$ encodes a change in the effective temperature of $f_0$,
and from Eq.~\eqref{eq:linearizing-collision-operator-2} we see that this implies that the eigenstate of $L$ with the lowest eigenvalue has eigenvalue $l=0$, which in turn means that the lowest eigenvalue of the late-time $H$
of Eq.~\eqref{eq:late-time-H} is zero. All the other eigenvalues of $H$ grow $\propto e^y D(y)$ at late times once the term \eqref{eq:late-time-H} dominates the Hamiltonian.
The lowest state, with zero eigenvalue according to \eqref{eq:late-time-H},
describes the evolution of the effective temperature, which is to say it describes the hydrodynamic evolution of the late-time solution around which we are expanding. 
The precise value of the lowest eigenvalue of the full $H$ given by \eqref{eq:H-DK-L} will be determined by the spectrum of $K$.
Note that all the  eigenvalues of $H$ other than the lowest control both how quickly the perturbations in the initial condition decay and
how quickly the contributions to $\Delta w$ coming from the source term $K w_0$ in \eqref{eq:Delta-w-EoM} decay. 

Therefore, we have proved what we set out to demonstrate: the eigenvalues of $H$ grow exponentially as $e^{(1-\delta)y}$, where $\delta \equiv -\partial_y D/D$.
In terms of $\tau$, we have demonstrated power law growth for 
all the eigenvalues of $H$ other than the lowest, and hence for the gap above the ground state of $H$, 
given by $\tau^{1-\delta}$.
In turn, this means
an \textit{exponentiated} power law %{\color{magenta} BSH: we may want to use a different description because what appears in the exponent is $\tau^{1-\delta}$, not just $\tau$} 
decay $e^{- \# \tau^{1-\delta}}$ for the occupation of all eigenstates other than the lowest eigenstate of $H$.  This makes the ``attractiveness'' of hydrodynamics manifest.

At late times after hydrodynamization,
the evolution will be described by ideal hydrodynamics as viscous effects become less and less important.  In a system
that is homogeneous in the transverse plane and that has boost-invariant longitudinal expansion, such as the one we have studied here, 
ideal hydrodynamics means $\delta = 1/3$ will take over (where $\partial_y \varepsilon/\varepsilon = -4/3$ and $\varepsilon \propto D^4$), and all the eigenvalues of $H$ other than the lowest will grow $\propto e^{(2/3)y} = \tau^{2/3}$, as evidenced at late times in Fig.~\ref{fig:elevels-Log}. In future work, it will be interesting to see the interplay between this phenomenon and the effects of inhomogeneities in the transverse plane, which can keep viscous corrections active for longer. We can also imagine (also for future work) investigating how the rapid growth $\propto \tau^{2/3}$ 
of the eigenvalues, and consequent decoupling, of every non-hydrodynamic mode
is reflected in the speed and manner with which a hydrodynamic fluid that has been perturbed by a passing jet rehydrodynamizes. Finally, it will also be important in future work to investigate how this analysis and these conclusions are modified when the longitudinal expansion is not strictly boost invariant. For example, if the fluid has some wide but finite extent in spacetime rapidity $\eta_s$ as in a heavy ion collision and the longitudinal expansion is approximately boost invariant only over some range $|\eta_s|<\eta_s^{\rm max}$, then at very late times when $y\gtrsim\eta_s^{\rm max}$ (which will be after the evolution has become that of ideal hydrodynamics)  
the prefactor in \eqref{eq:late-time-H} will grow more slowly than $e^y$ and our analysis will need to be revisited.

To summarize, in this Section we have shown that for a broad class of kinetic theories undergoing boost-invariant expansion, once the momentum distribution is close to isotropic the time dependence of the energy levels seen in Figure~\ref{fig:elevels-Log} is generic as and after the system hydrodynamizes.\footnote{It is tempting to think that we could have arrived at this conclusion by evaluating $H[w]$ as we introduced it earlier (without linearizing) on the thermal distribution and calculating its energy spectrum. 
This argument alone is incomplete, however: we would then have needed to justify why to evaluate $H$ on the equilibrium distribution state as opposed to the actual state of the system, which is to say we would then have needed to
assume that these two states are sufficiently similar.  This leads to linearizing the equations, as we did from the beginning in the actual derivation that we have presented.
} 
(This is so as long as the kinetic theory is such that the distribution will eventually isotropize and hydrodynamize, as long as there are no dimensionful scales other than $D(y)\sim T$, and as long as this scale drops more slowly than $1/\tau$.)
This has important consequences regarding the speed with which hydrodynamics is approached relative to the timescale of hydrodynamic evolution
---in other words, regarding how the non-hydrodynamic modes of kinetic theory decouple from the hydrodynamic sector. In non-expanding systems, it is questionable whether complete hydrodynamization ever takes place if the linear operator $L$ is gapless~\cite{Gavassino:2024rck}, which is to say if arbitrarily long-lived non-hydrodynamic modes are present, a situation that is not uncommon. The obstruction is that in such a situation there is no scale separation between the rates at which hydrodynamic and non-hydrodynamic degrees of freedom evolve. However, 
our results confirm that the conclusion is quite different in a boost-invariant expanding system.
We have shown that, in such a system, the exponentiated power law decay of any non-hydrodynamic mode will, after sufficient time, render such a mode unimportant relative to the power-law timescale of boost-invariant hydrodynamic evolution.

\section{Outlook} \label{sec:outlook}

We have shown that the AH framework is a natural and general way of simultaneously understanding both pre-thermal and hydrodynamizing attractor behavior using a simple kinetic theory that includes both inelastic and elastic processes and which captures many of the physical features expected of hydrodynamization in QCD EKT. Using AH, one can predict, understand, and obtain a quantitative description of, the emergence of attractor behavior and the nature of the decay of excited state modes (be it exponential or polynomial decay) by analyzing the spectrum of the effective Hamiltonian operator $H$, even in the absence of scaling behavior in the gluon distribution function. 

The AH framework 
makes the loss of memory that is the defining characteristic of hydrodynamization 
manifest, and allows us to see how it happens
in two sequential stages: first, via attraction to a pre-hydrodynamic attractor surface described by a band of ground states of $H$ separated from all other eigenstates (whose initial occupation has been 
forgotten); followed later by the opening up of a further gap above a unique ground state that is the hydrodynamizing attractor.  This sequential loss of memory via an attractor surface followed by a unique attractor is described equally naturally in the AH framework whether or not the kinetic theory exhibits scaling.
This provides a good example of how the AH framework expands the regime of applicability of 
the quantitative description and qualitative understanding of pre-thermal attractor behavior,
particularly to stronger couplings at which scaling breaks down.   We have also confirmed that, as first demonstrated in Ref.~\cite{Kurkela:2015qoa}, a kinetic theory that includes inelastic processes describes hydrodynamization occuring in a realistically rapid timescale when the coupling in the kinetic theory is increased to realistic values.

An aspect that is missing from our simplified kinetic theory is the Bose factors. Their absence introduces a qualitative difference in the behavior of the system at very early times when the occupancies are still large, and eliminates the possibility of early Bose condensation before number-nonconserving inelastic processes gain traction. The Bose enhancement of the effective density of scatterers has a well-understood effect on the eigenvalues of $H$ and was included in both Refs.~\cite{Brewer:2022vkq} and~\cite{Rajagopal:2024lou}, but the Bose enhancement of the non-diffusive piece of the $2\leftrightarrow 2$ kernel as well as the Bose-corrections to the  $1\leftrightarrow 2$ kernel have not been explored using the lens of AH. 
We have also neglected the possibility that plasma instabilities may play a role at very early times~\cite{Arnold:2003rq,Bodeker:2005nv,Kurkela:2011ub}. Addressing these omissions could allow us to employ the AH framework at earlier times than the $\tau_I$ at which we have begun our analyses in this paper, and would shed light on  
the state of the system at $\tau\sim \tau_I$ before it is attracted to the pre-hydrodynamic attractor.

In Section~\ref{sec:hydro_inevitability}, we have shown that the late-time behavior (during and after hydrodynamization) of the eigenvalues of $H$ that describe all the non-hydrodynamic modes that is apparent in our results is in fact much more general than the particular kinetic theory of Section~\ref{sec:results}. Assuming that the three broadly applicable criteria that we have specified are applicable (in essence, a kinetic theory of massless particles, with an isotropic solution, expanding such that the effective temperature drops more slowly than $1/\tau$),
after hydrodynamization has begun all the eigenvalues of $H$ other than the lowest (which describes the hydrodynamic evolution itself)
grow exponentially in $\ln \tau$, meaning as a power of $\tau$.  At late enough times that the evolution can be approximated by ideal hydrodynamics,
all non-hydrodynamic eigenvalues of $H$ grow like
$\tau^{2/3}$. This means that the occupation of all non-hydrodynamic modes decreases exponentially in a power of $\tau$ once hydrodynamization begins, with that power tending to $2/3$ at late times. This exponentiated power-law decay is much faster than the power-law timescales over which hydrodynamic quantities evolve.  
(This conclusion applies whether the time-independent collision kernel operator is gapless or gapped.)  
So, at later and later times after hydrodynamization has begun, the hydrodynamic attractor becomes more and more attractive. 
These conclusions have been derived in the absence of transverse spatial gradients.  We leave the investigation of how robust they are to spatial gradients to future work. This investigation is a necessary precursor to drawing conclusions from the 
rapid power-law growth of all non-hydrodynamic eigenvalues and the even more rapid exponentiated power law decoupling (decay of their occupations)
for how, and how rapidly, a hydrodynamic fluid that has been perturbed by a passing jet rehydrodynamizes.

We look ahead with considerable anticipation to forthcoming work incorporating transverse spatial gradients into kinetic theory in the AH framework~\cite{BrewerSharell}, as this may open the door to implementing AH in settings with more realistic (in particular, not homogeneous) transverse profiles. A combination of the simple but reasonably physical number-non-conserving kinetic theory that we have employed here with non-trivial transverse spatial gradients may be a good recipe for a practical AH description of the weakly coupled kinetic theory dynamics throughout the pre-hydrodynamic stage in a simulation of heavy ion collisions.

\acknowledgments

We are grateful for helpful conversations with Jasmine Brewer, Lorenzo Gavassino, Jamie Karthein, Aleksi Kurkela, Govert Nijs, Jean-Fran\c{c}ois Paquet, Uri Sharell, and Gabriel Soares Rocha.
This research was supported in part by the U.S.~Department of Energy, Office of Science, Office of Nuclear Physics under grant Contract Number DE-SC0011090, by grant NSF PHY-2309135 to the Kavli Institute for Theoretical Physics (KITP), and by grant 994312 from the Simons Foundation. KR acknowledges the hospitality of the CERN Theory Department.

\bibliography{main.bib}

\end{document}